\documentclass[prb,aps,superscriptaddress,amsmath,amssymb,floatfix,twocolumn,showpacs,amsfonts,longbibliography]{revtex4-1}
\usepackage{bm}
\usepackage{mathrsfs}

\usepackage{textcomp}
\usepackage{graphicx}
\usepackage{subfigure}

\usepackage{tabularx} 
\usepackage{color}
\usepackage{braket}
\usepackage{overpic}
\usepackage{gensymb}
\usepackage{tikz}
\usepackage{multirow}
\usepackage{paralist}
\usepackage{indentfirst}
\usepackage{soul}
\usepackage{cancel}

\allowdisplaybreaks

\usepackage[colorlinks=true,citecolor=blue,urlcolor=blue,linkcolor=blue,hyperindex]{hyperref}

\begin{document}
\title{Loop Algorithm for Quantum Transverse Ising Model in a Longitudinal Field}
\author{Wei Xu}
\affiliation{Department of Physics, and Chongqing Key Laboratory for Strongly Coupled Physics, Chongqing University, Chongqing, 401331, China}

\author{Xue-Feng Zhang}
\thanks{corresponding author:  zhangxf@cqu.edu.cn}
\affiliation{Department of Physics, and Chongqing Key Laboratory for Strongly Coupled Physics, Chongqing University, Chongqing, 401331, China}
\affiliation{Center of Quantum Materials and Devices, Chongqing University, Chongqing 401331, China}

\begin{abstract}
The quantum transverse Ising model and its extensions play a critical role in various fields, such as statistical physics, quantum magnetism, quantum simulations, and mathematical physics. Although it does not suffer from the sign problem in most cases, the corresponding quantum Monte Carlo algorithm performs inefficiently, especially at a large longitudinal field. The main hindrance is the lack of loop update method which can strongly decrease the auto-correlation between Monte Carlo steps. Here, we successfully develop a loop algorithm with a novel merge-unmerge process. It demonstrates a great advantage over the state-of-the-art algorithm when implementing it to simulate the Rydberg atom chain and Kagome qubit ice. This advanced algorithm suits various systems such as Rydberg atom arrays, trapped ions, quantum materials, and quantum annealers.
\end{abstract}
\maketitle

\section{introduction}
In the tapestry of statistical physics, the quantum transverse Ising model (QTIM) is a thread of profound significance, weaving together the principles of quantum mechanics with the stochastic nature of thermal fluctuations. It serves as a canonical model for understanding phase transitions and critical phenomena, particularly those of a quantum nature \cite{QTIM_1d}. The interplay between geometry frustration \cite{frus_ising_1,frus_ising_2}, long-range interactions \cite{Rydberg_longr}, and many-body effects \cite{fracton01} leads to a multitude of complex and exotic phenomena, including fractionalization \cite{frus_ising_3}, emergent lattice gauge theory \cite{lgt_review,lgt_changle,Rydberg_gauge}, glass states \cite{Rydberg_glass}, and fracton excitations \cite{fracton02}. Additionally, the integrability of these systems has garnered significant interest in the field of mathematical physics \cite{Baxter}.

Experimentally, the family of QTIM is instrumental in analyzing the quantum phase and phase transition within quantum materials, ranging from frustrated magnetism \cite{tmgo_kt,tmgo_model,tmgo_neu,tmgo_uud,trap_ion} to paraelectric hexaferrite \cite{chai}. Furthermore, it can also well describe cutting-edge quantum many-body simulators, such as the Rydberg atom arrays \cite{Rydberg_chain,Rydberg_nature1,Rydberg_nature2}, trapped ions \cite{trap_ion}, and commercial quantum annealers like D-WAVE \cite{dwave,Kagome_dwave,kagome_dwave2}. Consequently, in both theoretical and experimental aspects, the development of numerical simulation techniques has become crucial and pressing.

Due to the absence of the sign problem in most cases, the best choice would be the quantum Monte Carlo (QMC) method. To eliminate the discretization error associated with the Trotter-Suzuki decomposition and fix the problem of the directed loop algorithm due to the lack of XY interaction\cite{SSE_Sandvik_2, Syljuåsen_2003}, A. W. Sandvik developed a stochastic series expansion (SSE) QMC algorithm specifically tailored for the QTIM \cite{SandVik_SSEforQIM}. However, the Swendsen-Wang cluster-type update process within the algorithm becomes inefficient when including the longitudinal field {which becomes highly relevant to the exotic quantum phases in the quantum frustrated magnetism \cite{PhysRevLett.84.4457,pnas.2015785118,Rydberg_glass,PhysRevX.15.011025,PhysRevB.103.104416,bombieri2025deconfinedquantumcriticalitytriangular,Ishizuka_2011}. } Although the catastrophe of ultra-low acceptability rates can be strongly relieved by subdividing each cluster into lines \cite{Melko_SSEforRydberg} or designing a more complex scheme \cite{QMC_rydberg}, the most effective solution is expected to be the loop \cite{qmc_loop} or worm algorithm \cite{worm1,worm2}. However, lacking spin exchange interactions makes the loop algorithm of QTIM a long-standing problem \cite{SSE_Sandvik_1,SSE_Sandvik_2}.

\begin{figure}[t]
	\centering
	\includegraphics[width=0.99\linewidth]{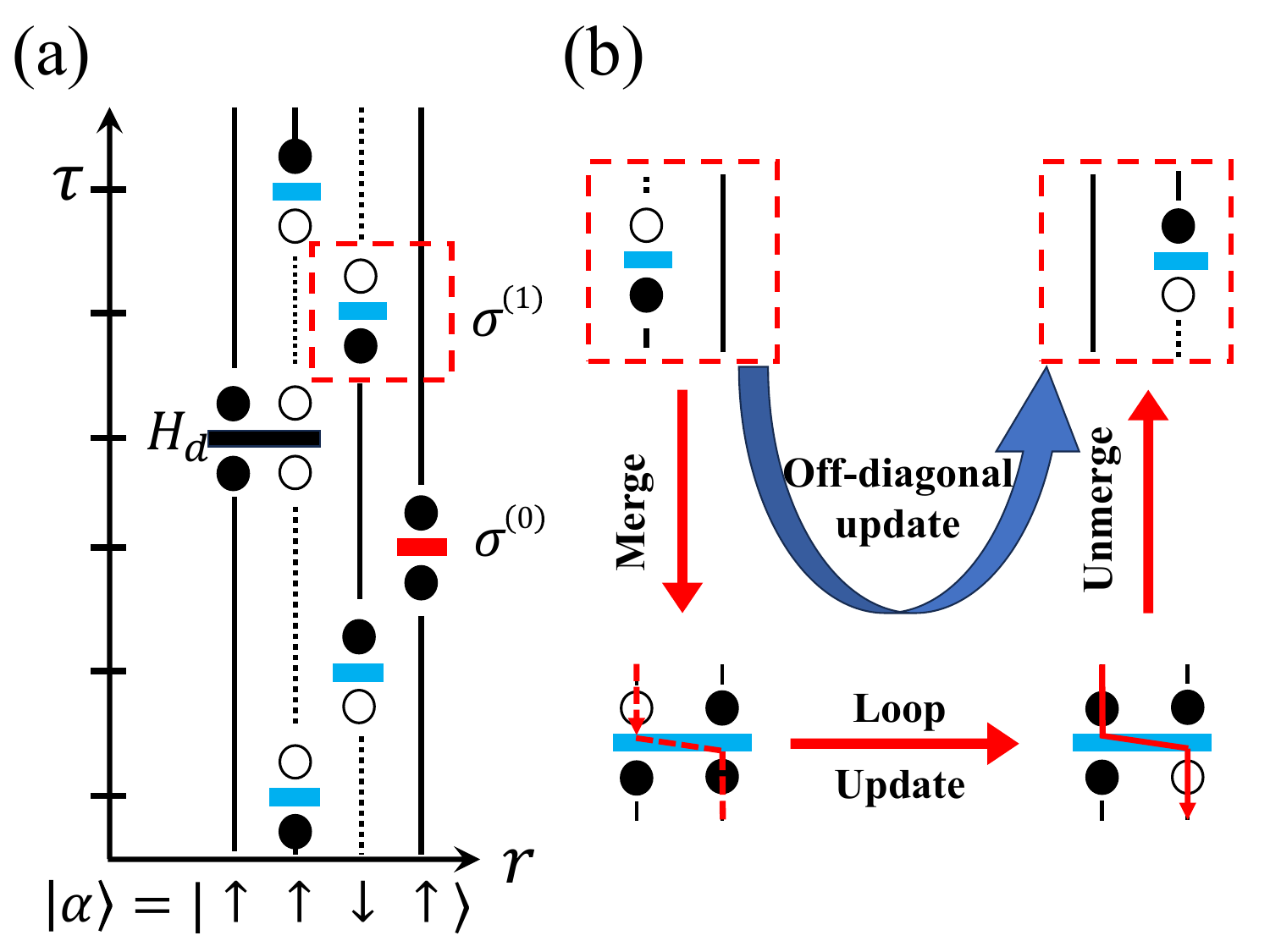}
	\caption{Schematic diagram of the loop algorithm. (a) During the loop update process, the QMC sample is depicted in the diagram with the diagonal operator $H_d$ (black bond), constant operator $\sigma^{(0)}$ (red bond), and off-diagonal operator $\sigma^{(1)}$ (blue bond) locating at each integer imaginary time. The black (white) circles denote the spin up (down). (b) The merge-unmerge processes can move the off-diagonal operator to the other site, in conjunction with the loop update process which flips the spins along the path.
 }\label{fig1}
\end{figure}

In this manuscript, a novel update strategy of the loop algorithm is designed to simulate the QTIM family with enhanced efficiency. As shown in Fig.~\ref{fig1}, by inventing the merge-unmerge update processes, the worm can move to the other site so that the position of the off-diagonal operator can be altered in the spatial dimension. Then, we implement the algorithm on a realistic system: Rydberg atom chain and challenging frustrated system: Kagome qubit ice, respectively \cite{sup}. Compared to the conventional directed loop algorithm \cite{Syljuåsen_2003} and current state-of-the-art line algorithm \cite{Melko_SSEforRydberg}, our results demonstrate that the loop algorithm exhibits much shorter auto-correlation times, with its advantages becoming more pronounced in the presence of large longitudinal fields.

\section{loop algorithm}
\subsection{SSE framework}
{We begin with a brief review to the SSE method. The partition function is expanded as a power series:
\begin{equation}
    Z=\sum_{\alpha} \sum_{n=0}^{\infty} \frac{\beta^{n}}{n!} \bra{\alpha} (-H)^n \ket{\alpha}, \label{series}
\end{equation}
where $\beta=1/T$ is the inverse of the temperature, $\ket{\alpha}$ is the spin basis, a basis of the Hilbert space and $H$ is the Hamiltonian simulated. Specifically in this manuscript, 
\begin{equation}
    H=H_d-\Gamma \sum_i \sigma_i^{(1)} -\Gamma \sum_i \sigma^{(0)}_i, \label{Hamilton}
\end{equation}
where $\sigma^{(k)}$ represents the Pauli matrix, $H_d=J_z\sum_{i,j} \sigma_i^{(3)}\sigma_i^{(3)}-B\sum_{i} \sigma_i^{(3)}$ is the diagonal operator that encompasses Ising interaction ($J_z$) and longitudinal field ($B$), $\Gamma$ characterizes the strength of the transverse field or off-diagonal operator, and the last energy shift term is introduced to transfer with off-diagonal operator and referred to as constant operator. 

Note that each term in Eq.~\ref{Hamilton} maps a basis state $\ket\alpha$ to a state $\ket{\alpha^\prime}$ within the same basis, so $\bra{\alpha}(-H_{v_u})\ket{\alpha^\prime}$ is a real number. As a result, we can decompose $(-H)^n$ in Eq.~\ref{series}, 

\begin{align}
    Z = & \sum_{n=0}^{\infty} \frac{\beta^{n}}{n!}
         \sum_{\alpha_1}\sum_{v_1,v_2,...,v_n} \bigg\{ \nonumber \bra{\alpha_1} (-H_{v_1}) \ket{\alpha_{2}}
         \\
         & 
          \bra{\alpha_2} (-H_{v_2}) \ket{\alpha_{3}}
         \cdots
         \bra{\alpha_{n}} (-H_{v_n}) \ket{\alpha_{1}} \bigg\},
    \label{decomposition}
\end{align}

for the Monte Carlo sampling. In Eq.~\ref{decomposition}, $H_{v_u}$ corresponds to the $v$th term in Eq.~\ref{Hamilton} of $u$th $H$ in Eq.~\ref{series}. The partition function is thus expressed as a sum over all possible operator sequences depicted diagrammatically in Fig.~\ref{fig1} (a), where a black bond represents $H_{v_u} = H_d$, a blue bond represents $H_{v_u} = -\Gamma \sigma_i^{(1)}$ and a red bond represents $H_{v_u} = -\Gamma \sigma_i^{(0)}$.

The QMC update algorithm typically consists of diagonal and off-diagonal parts\cite{SSE_Sandvik_1, SSE_Sandvik_2}. The diagonal update stochastically inserts and removes diagonal operators (including the constant operator) at various positions (imaginary time slices) along the operator sequence. 

To perform the diagonal update,  the partition function should be  reformulated by truncating the series expansion at a maximum power equals $M$ and inserting $ M - n$ identity operators($\mathbb{I}$). There are $C_M^n$ ways to insert these identities, leading to:

\begin{align}
    Z^\prime = & \sum_{n=0}^{M} \frac{\beta^{n}(M-n)!}{M!}
         \sum_{\alpha_1}\sum_{v_1,v_2,...,v_n} \bigg\{ \nonumber \\
         & \bra{\alpha_1} H^\prime_{v_1} \ket{\alpha_{2}}
         \bra{\alpha_2} H^\prime_{v_2} \ket{\alpha_{3}}
         \cdots
         \bra{\alpha_{n}} H^\prime_{v_n} \ket{\alpha_{1}} \bigg\},
    \label{truncated_}
\end{align}

in which $H^\prime_{v_u} = -H_{v_u}$ or $\mathbb{I}$. Then, the diagonal update subsequently checks and tries to exchange every operator that is not an off-diagonal operator(which would change the configuration) with the identity operator by the Metropolis probability:
\begin{equation}
    P(\mathbb{I}\to H_{v_u})=\min\left(1, \frac{N\beta \langle \alpha | (-H_{v_u}) | \alpha \rangle}{M - n}\right),
    \label{insertion}
\end{equation}
\begin{equation}
    P(H_{v_u} \to \mathbb{I})=\min\left(1, \frac{M - n + 1}{N\beta \langle \alpha | (-H_{v_u}) | \alpha \rangle}\right),
    \label{removal}
\end{equation}

where $N$ is the number of possible operator types, insertion means replace $\mathbb{I}$ by $-H_{v_u}$ and removal is the reverse process.  Notice that, there is no system error introduced by the truncation of the operators because $M$ is adjusted to a large enough value $n$  never could reach within finite simulation time.

}

In comparison, the off-diagonal update focuses on swapping the constant and off-diagonal operators to ensure the ergodicity of QMC. In conventional methods\cite{SandVik_SSEforQIM, Melko_SSEforRydberg}, the diagram can be divided into many segments with constant and off-diagonal operators serving as boundaries. Subsequently, off-diagonal updates are executed by flipping spins within randomly selected clusters. However, the acceptance rate significantly decreases when dealing with large clusters in the presence of a substantial longitudinal field. Drawing from the historical development of QMC methods \cite{qmc_loop}, it is logical to anticipate that a loop algorithm could address these challenges.

we have added dashed-line boxes to highlight the evolution of operators during the merge-unmerge steps

\subsection{Loop update}
\begin{figure}[t]
	\centering
	\includegraphics[width=0.99\linewidth]{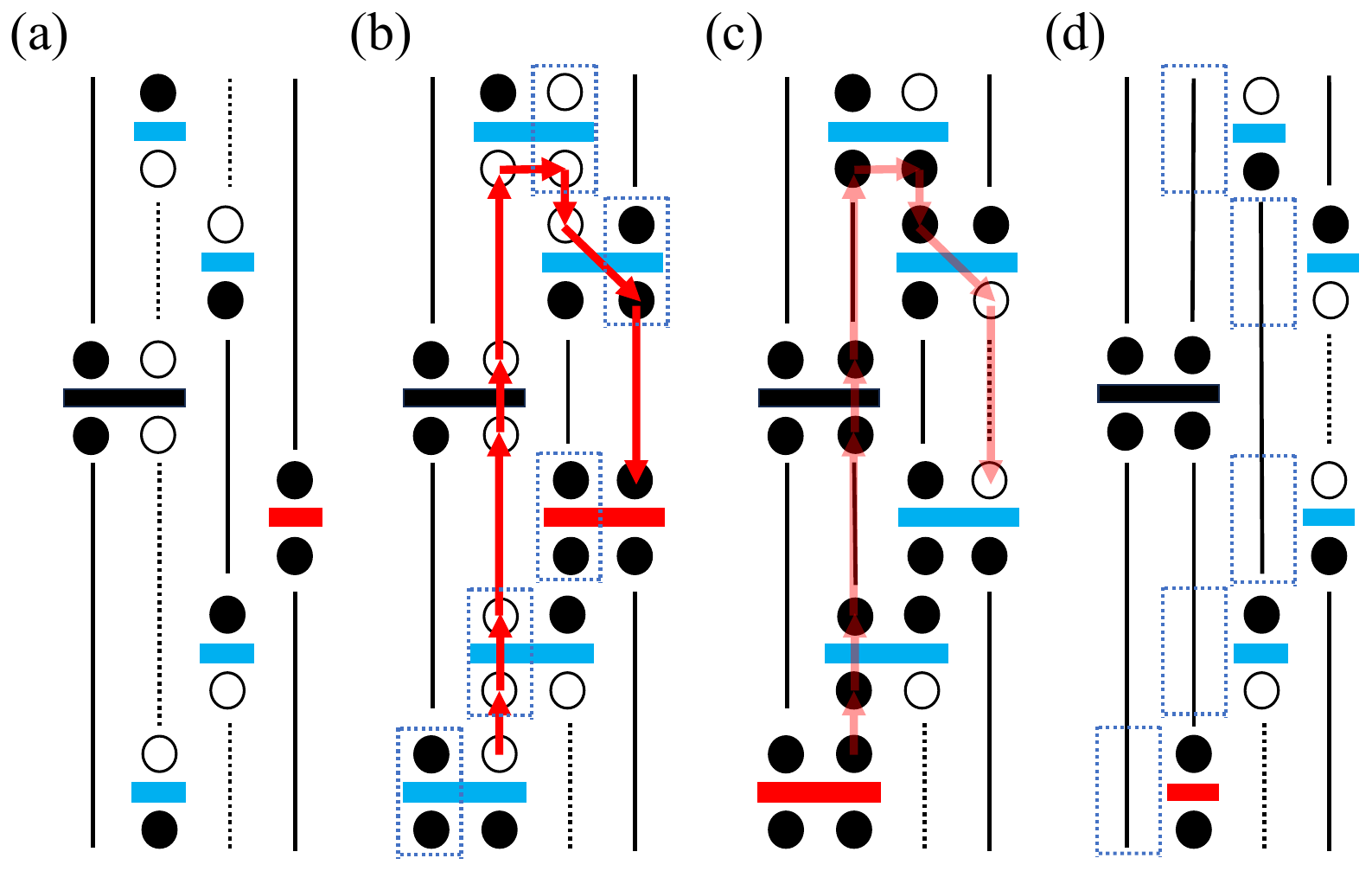}
	\caption{Schematic picture of the off-diagonal update process: (a) The initial configuration. (b) The loop path after the merge step; the dashed boxes indicate the merged sites. (c) The resulting configuration after flipping all spins along the loop path. (d) The final configuration after the unmerge process; the dashed boxes now indicate the unmerged sites.
 }\label{fig2}
\end{figure}

Our off-diagonal update process is designed to involve three distinct stages:\textbf{ merge, loop, and unmerge}.{Fig.~\ref{fig2}  illustrates the off-diagonal update process.} As shown in Fig.~\ref{fig1} (b) and Fig.~\ref{fig2}(a)(b), at the beginning of the off-diagonal update, the single-site operator (constant and off-diagonal operators) can merge with a random site to construct a merged operator. {The dashed-line boxes highlight the evolution of operators during the merge-unmerge steps.} For the sake of simplicity, only the nearest neighbor sites are considered for this merging process. Then, at the loop stage, a worm (also called the loop head) is created at a leg. As it moves, it flips the spin configuration at every site it traverses. At last, with the help of the unmerge process, the merged operator will transfer back to the single-site operator, so that the off-diagonal operator can move to different sites.

{For the loop stage,} the worm here does not necessitate forming a closed loop because the magnetization is not conserved due to the transverse field. Consequently, the worm undergoes the \textbf{ start-run-stop procedure}.
In the start process, a merged operator is randomly selected. For the constant merged operators, any of the four legs can serve as the starting point. In contrast, for the off-diagonal merged operators, only two legs are eligible for selection; choosing otherwise would result in the emergence of an invalid operator. Then, the spin at the initial leg is flipped, and these two distinct types of merged operators can interchange with each other.

When the worm runs on the configuration, it will meet three different operators, and corresponding transfer strategies are different as illustrated in Fig.~\ref{fig2}(b). (i) \textbf{Diagonal operator}: Only direct passing through and bounce-back are allowed because lack of spin exchange operators, and the acceptability follows the Metropolis way. (ii) \textbf{Off-diagonal merged operator}: The worm can randomly exit at one of the other three legs with equal probability. (iii) \textbf{Constant merged operator}:  The worm always passes through directly. All spins along the path are flipped except for the bounce-back process since the entrance leg is visited twice. 

As an open loop, the worm has the flexibility to terminate at any merged operator. To tune the length of the loop, we introduce a free parameter named the loop-stop probability $P_s$. When $P_s$ equals one, the worm will immediately stop upon encountering the first merged operator, and effectively the loop algorithm turns back to line algorithm \cite{Melko_SSEforRydberg}. On the other hand, the loop will never stop at $P_s$=0.  To avoid introducing any bias, we set $P_s=1/2$ here. The discussion about $P_s$ can be found in the Appendix \ref{app:stop_p}. When the worm meets an off-diagonal merged operator, it can only stop at the legs with different states.  In contrast, any leg of the constant merged operator can be chosen as the ending point but with an acceptance probability of $P_s/2$. This reduced probability is because the worm has only a half chance of encountering the constant operator at the correct position to stop. The discussion about the loop length can be found in Appendix \ref{app:loop}.

To preserve complexity, the number of loops in each QMC step is set to ensure each operator can be visited at least twice on average, a rule also followed by the line algorithm. After these loops are finished, the final unmerge step has to be executed, see Fig.~\ref{fig2}(c). For the off-diagonal operator, the unmerge process involves retaining one side with distinct spin states, e.g. $\frac{\circ\bullet}{\bullet\bullet}\longrightarrow\frac{\circ}{\bullet}$. The constant merged operator would randomly keep one side to be consistent with the 1/2 in loop-stop probability $P_s/2$ and the equal probability to choose the start leg of the constant operator, which is the requirement of the detailed balance. After that, the entire off-diagonal update process is completed, and the measurement part is the same as the conventional algorithm \cite{measure}. The discussion about the detailed balance condition can be found in the Appendix A.

To demonstrate the advantages of the loop algorithm, we compare it with the conventional loop algorithm \cite{Syljuåsen_2003} (with setting all XY-term coefficients to zero and labeled as $\mathrm{loop}_c$) and the state-of-the-art line cluster algorithm in two typical systems: the Rydberg atom chain and the Kagome qubit ice. To make sure the same complexities of both algorithms, we use the same codes except for the off-diagonal update implemented with different updates, respectively.  The best quantity to qualify the efficiency of the QMC algorithm is the auto-correlation function  $C(t) = \langle O_{i}O_{i+t}\rangle - \langle O_{i}\rangle\langle O_{i+t}\rangle$, in which $t$ denotes the QMC steps, and $O_i$ is the observable calculated in $i$th sample \cite{qmc_loop}. It usually follows the exponential decay behavior and the auto-correlation time $\tau_{\mathrm{mc}}$ is defined as the time $C(t)$ drops to $1/e$ of $C(0)$ so that two QMC samples with a distance of $\tau_{\mathrm{mc}}$ in the sequence can be considered as independent. In the following simulations, the auto-correlation time is calculated with the integrated method \cite{qmc_loop} from $10^7$ successive Monte-Carlo measurements or the Markov chain process. Then, its error is obtained by taking sixty such independent Markov chain processes. The comparison of the algorithm has to be implemented in the same computing environment, e.g. CPU, memory, program language, and compiler.

\begin{figure}[t]
	\centering
	\includegraphics[width=0.99\linewidth]{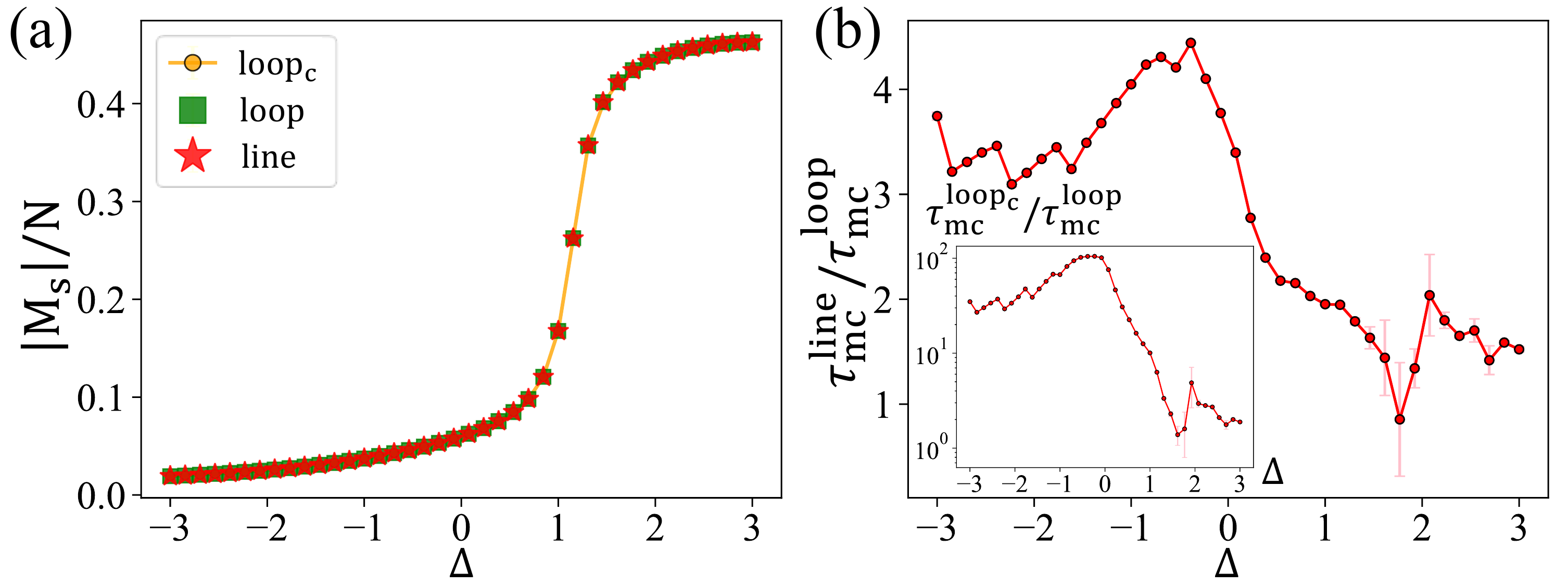}
	\caption{\textbf{Rydberg atom chain.} \textbf{(a)} Staggered magnetization per site $\mathrm{|M_s|/N}$ and \textbf{(b)} the ratio of its auto-correlation time by line and loop algorithm (Inset: by $\mathrm{loop}_c$ and loop algorithm).
 }\label{fig3}
\end{figure}

\section{results}
\subsection{Rydberg Atom Chain}
The Rydberg atom array provides a highly tunable platform to simulate quantum magnetism \cite{browaeys2020many}. In the experiment, many ultra-cold two-level atoms are individually trapped by the tweezer lights. Thanks to the rapid development of the experimental technique, each tweezer light is independently tunable, including but not limited to its strength, position, and frequency. Therefore, various geometry can be realized, such as chain \cite{Rydberg_chain}, square \cite{Rydberg_nature1, Rydberg_nature2}, and triangular lattice \cite{Rydberg_nature2}. Meanwhile, the Rydberg state of the atom can be excited by the two-photon process so that the strong interaction between Rydberg atoms can be introduced. The corresponding Hamiltonian can be expressed as follows:
\begin{equation}
H = \sum_{i<j} V_{ij} n_in_j-\Delta\sum_i n_i-\frac{\Omega}{2}\sum_i  \sigma^{(1)}_i, \label{E2}
\end{equation}
where $\sigma^{(1)}_i=\ket{g}_i\bra{r}_i+\ket{r}_i\bra{g}_i$ describes the excitation from the ground state $\ket{g}$ to the Rydberg state $\ket{r}$ at $i$th trap site with Rabi frequency $\Omega$, $n_i=\ket{r}_i\bra{r}_i$ is the density operator of atom in Rydberg state, $\Delta$ denotes the strength of the detuning and $V_{ij}=V/R_{ij}^6$ takes the form of repulsive van der Waals interactions. After the transformation $n_i\leftrightarrow\frac{\sigma^{(3)}_i+1}2$ and limiting the interaction to the nearest neighbor, we can find that the Rydberg atom chain model Eq.\ref{E2} can be mapped into the QTIM with correspondence: $J_z=V/4$, $B=(\Delta-V)/2$, and $\Gamma=\Omega/2$. Therefore, the loop algorithm is straightforwardly adaptable to the simulation of the Rydberg atom array system. The rapid decay of the interaction allows for an efficient simulation by including only interactions up to the third nearest neighbors.

When the repulsive interaction dominates, the atoms in the blockade radius $R_b=(V/\Omega)^{1/6}$ of the Rydberg atom can not be excited to the Rydberg state so that the system enters into the $\mathbb{Z}_2$ ordered phase characterized by the order parameter: staggered magnetization $|M_s|=|\sum_i(-1)^i(n_i-1/2)|$ \cite{Melko_SSEforRydberg,Rydberg_chain}. To simulate the quantum phase transition (QPT) from the ordered phase to the disordered phase, we set $R_b=1.2$ and inverse temperature $\beta=20$ with $\Omega = 1$ as the energy unit. The long-range interaction is truncated to the third nearest-neighbor site. The chain contains 51 sites with open boundary conditions \cite{Melko_SSEforRydberg,Rydberg_chain}.

As demonstrated in Fig.~\ref{fig3}(a), the staggered magnetization per site $|M_s|/N$ increasing from zero to a finite value indicates the QPT exists at $\Delta\approx\Omega$ and the identity of the simulation result from all three algorithms proves the correctness of the loop algorithm. Comparing the auto-correlation times of line $\tau_{\mathrm{mc}}^{\mathrm{line}}$ and loop algorithm $\tau_{\mathrm{mc}}^{\mathrm{loop}}$, we can find their ratio $\tau_{\mathrm{mc}}^{\mathrm{line}}/\tau_{\mathrm{mc}}^{\mathrm{loop}}$ (see Fig.~\ref{fig3}(b)) indicates the loop algorithm has a shorter auto-correlation time, in other words, faster to achieve same simulation precision. In the small longitudinal field region $1\lesssim\Delta\lesssim3$, the ratio is close to two. The large fluctuation around the quantum critical point may result from the critical slowing down. At a large longitudinal field where the line algorithm suffers from low acceptability, the loop algorithm demonstrates a significant improvement, offering approximately fourfold or even greater acceleration. The inset of Fig.~\ref{fig3}(b) demonstrates the comparison between the auto-correlation times of conventional loop  $\tau_{\mathrm{mc}}^{\mathrm{loop_c}}$ and loop algorithm $\tau_{\mathrm{mc}}^{\mathrm{loop}}$. When closing to the particle-hole symmetry point, the improvement of our algorithm on the auto-correlation time is not too much, and its advantage is most pronounced for $\Delta<1$, where it outperforms the conventional algorithm by more than an order of magnitude.

On the other hand, the design of the numerical algorithm not only affects the auto-correlation time but also strongly changes the real computation time \cite{RNG}. Although here we set both algorithms to have the same program complexity, the line algorithm has an additional cluster searching process ( commonly existing in the Swendsen-Wang algorithm ) which is expected more time-consuming than the merge-unmerge process in the loop algorithm. We record the computation time of diagonal and off-diagonal updates in both algorithms to check it. It is not surprising the diagonal part consumes almost the same time, because the codes of both algorithms are identical only except for the off-diagonal part. Then, we define the dimensionless computation time $T_C$ as the computation time of the off-diagonal part divided by the diagonal part. We found that the line algorithm spends a longer time which confirms our suspicion. In addition, the disadvantage of the line algorithm becomes serious at large negative detuning (longitudinal field), and it may be due to the formation of the large line cluster in the disordered phase.


\subsection{Kagome Qubit Ice}
Constructed with the corner shared triangle, the Kagome lattice exhibits strong geometry frustration so that the local Ising interaction leads to the ground state with macroscopic degeneracy following the ice rule \cite{qtim_kagome1}. After turning on both transverse and longitudinal fields, the degeneracy will be lifted and the ground state will change to the valence bond solid (VBS) with spontaneous translational symmetry breaking. The presence of the VBS phase is due to the so-called order-by-disorder mechanism and related to the emergent lattice gauge theory with fractional excitations \cite{qtim_kagome2,Rydberg_gauge,Kagome_dwave,kagome_dwave2}. The QTIM in the Kagome lattice has been realized in the D-WAVE platform where named Kagome qubit ice \cite{Kagome_dwave,kagome_dwave2}, and we take it as our final challenging test. 

\begin{figure}[t]
	\centering
	\includegraphics[width=0.99\linewidth]{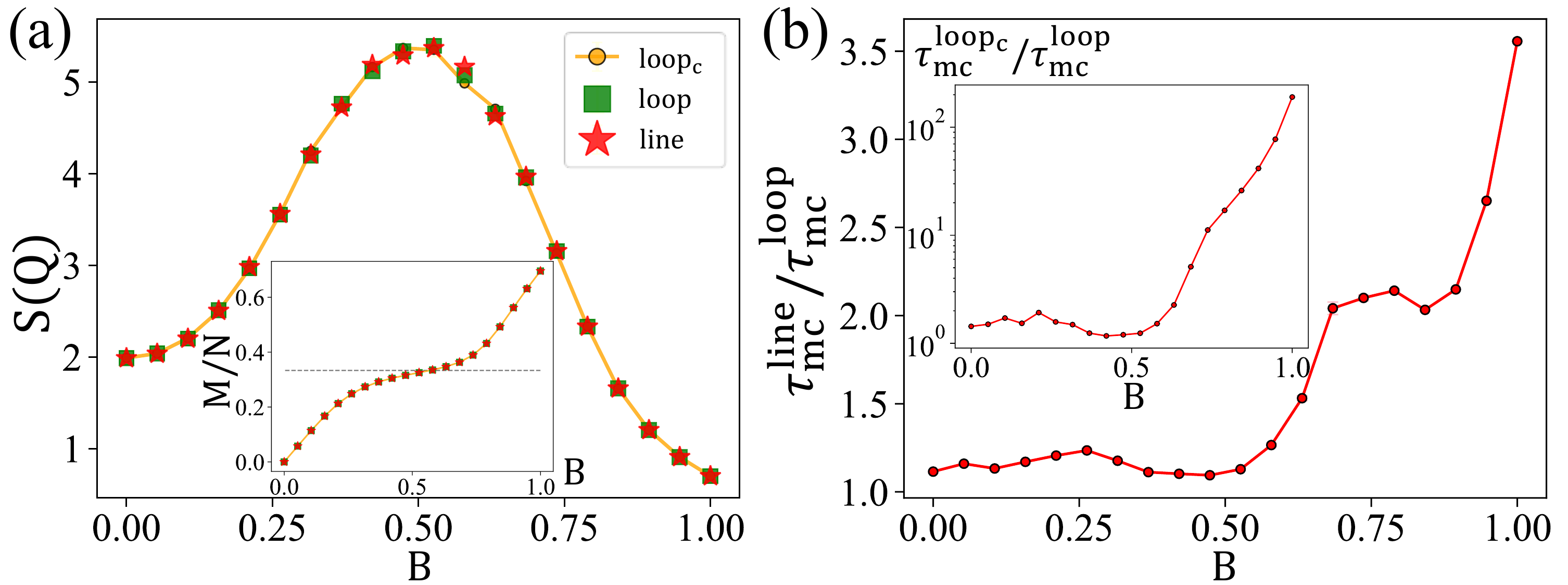}
	\caption{\textbf{Kagome qubit ice.} \textbf{(a)} Structure factor and \textbf{(b)} the ratio of its auto-correlation time by line and loop algorithm. The inset of (a) shows the magnetization per site and the dashed black line highlights the 1/3 corresponding to the ice rule filling. The inset of (b) shows the ratio of its auto-correlation time by $\mathrm{loop_c}$ and loop algorithm. The parameters are $J_z=0.25$, $\Gamma=0.15$, $\beta=20$, and $N=24\times24\times3$ with periodical boundary condition.
 }\label{fig4}
\end{figure}

The order parameter to describe the VBS phase is the structure factor defined as $S(\mathrm{Q})=\sum_{i,j}e^{i\mathrm{Q}r_{ij}}\sigma^{(3)}_i\sigma^{(3)}_j/N$ where $\mathrm{Q}=(\frac{4\pi}{3},0)$. As shown in Fig.~\ref{fig4}, the structure factors $S(\mathrm{Q})$ calculated by both algorithms match well and reach the maximum value around magnetization per site $M/N=\sum_i\sigma^{(3)}_i/N$ equals to 1/3, which indicates the formation of the VBS phase. Different from the XXZ model in the Kagome lattice \cite{kagome_zhang1,kagome_zhang2,kagome_zhang3}, the 1/3 magnetization plateau is not flat due to the nonconservation of the magnetization. The ratio of the auto-correlation times in Fig.~\ref{fig4}(b) strongly supports the advantages of the loop algorithm at the large longitudinal field. {Consistent with observations in Rydberg atom arrays, the loop algorithm demonstrates superior performance compared to both the line algorithm and conventional loop algorithms across the entire parameter range investigated. Specifically, for $B>0.5$, the auto-correlation time of the loop algorithm is 2-3 times shorter than that of the line algorithm. When compared to conventional loop algorithms, the reduction in auto-correlation time reaches up to two orders of magnitude, with this advantage becoming increasingly pronounced at higher longitudinal field strengths.} However, compared with the Rydberg atom chain, such advantages become not obvious in the small longitudinal field and may be due to the high degeneracy of the ground and excited quantum states. While introducing the {second and third nearest neighbor interactions}, our recent work demonstrates that geometric string breaking can be observed by tailoring the open edge in the Kagome Rydberg arrays \cite{xu2025geometricbreakingquantumstrings}.

\section{Conclusion and discussion}
By innovatively inventing a novel update strategy: merge-unmerge process, we successfully design a loop-type QMC algorithm tailored for the QTIM and its extensions. The merge-unmerge process can cause the loop update path to move in the spatial direction in the extended configuration space, while also avoiding the decrease in operator weight that occurs due to subdivision in the directed loop algorithm. After comparing with the state-of-the-art line cluster method by simulating two typical realistic platforms, the loop algorithm demonstrates significant advantages, especially in the large longitudinal field region. Meanwhile, the loop algorithm is more coding-friendly and the teaching code is open-source \cite{sup}. 

The loop algorithm can be viewed not only as an advanced numerical simulator for quantum magnetism, trapped ions, Rydberg atom arrays, quantum computers, and many other quantum many-body systems, but also easily extended to more complex models in statistical physics, such as the quantum clock and Potts models. Furthermore, such a merge-unmerge update process can also be ported to other Monte-Carlo methods, such as the continuous-time worm algorithm \cite{chunjiong}. Meanwhile, many advanced techniques related to the entanglement measurement \cite{Yan_1,Yan_2,Yan_3,Yan_4,Yan_5} can also be incorporated into our algorithm.

\section{Acknowledgment}
X.-F. Z. thanks Zheng Yan for fruitful discussions, MPI-PKS for hospitality during the visit, and acknowledges funding from the National Science Foundation of China under Grants  No. 12274046, No. 11874094, No.12147102, and No.12347101, Chongqing Natural Science Foundation under Grants No. CSTB2022NSCQ-JQX0018, Fundamental Research Funds for the Central Universities Grant No. 2021CDJZYJH-003, and Xiaomi Foundation / Xiaomi Young Talents Program.

\appendix
\section{Detailed Balance Condition}
\label{app:balance}

The detailed balance condition of the diagonal update is the same as Ref. \cite{PhysRevB.30.1599}. For the loop to pass through the operators, the updating methods are conventional, so the detailed balanced condition is also kept. Therefore, the merge-unmerge process and the starting and stopping of the loop processes should be discussed in detail.

The merge-unmerge process is similar to the link-cluster process in the Swendsen-Wang algorithm and the start and stop processes of the loop update in the directed loop algorithm. The balance condition should be satisfied between the original configuration space and the extended configuration space including merged operators. Because the unmerge process is the exact reverse process of the merge process, the detailed balance condition is kept. 

In the loop updating process, the detailed balance can be verified by analyzing a loop update path with its inverse path. Because the number of the merged operators $N_o$ is not changed, the probability $1/N_o$ due to random selection of the starting position of the loop is equal for all the loops in each loop update. Therefore, it is not necessary to include it into the acceptability of starting or stopping the loop. Then, the possibility that the loop stops after passing the $N_s$ merged operator is $(1-P_s)^{N_s}P_s$. Because the loop and its reverse process pass the same number of merged operators, this possibility related to the stopping process will not affect the detailed balanced conditions.  

\begin{figure}[t]
	\centering
    \includegraphics[width=0.95\columnwidth]{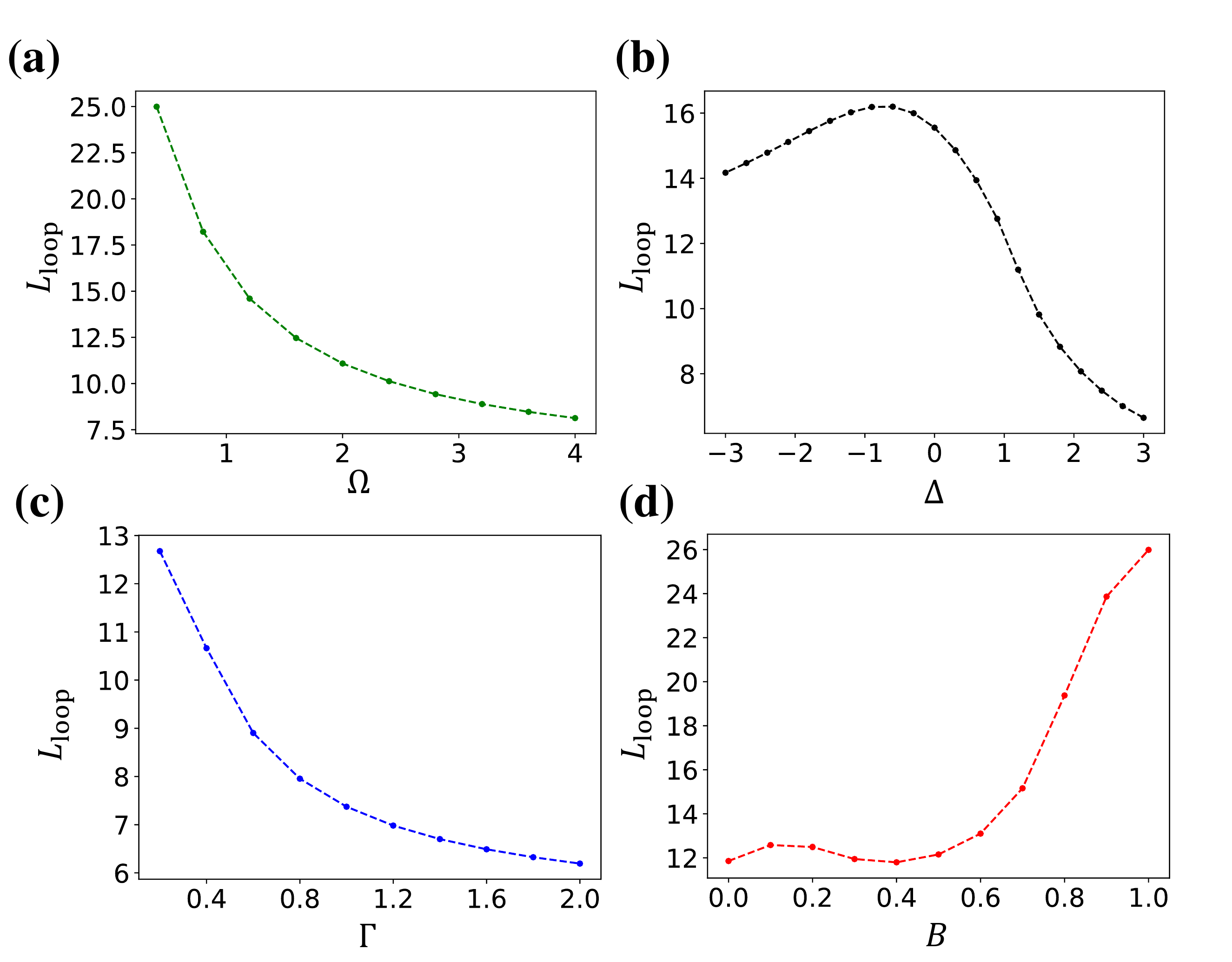}
	\caption{\label{replyfig2} The loop length of (a,b) the Rydberg atom chain and (c,d) Kagome qubit ice. (a) The Rabi frequency $\Omega$ is changing at $\Delta =0 $ and (b) the detuning $\Delta$ is changing at $\Omega=1.0$. The other parameters are identical to those in Fig.~\ref{fig3}: $R_b=1.2$, $\beta=20$, and $L=51$. (c) The transverse field $\Gamma$ is changing at $B$=0.5 and (d) the longitudinal field $B$ is changing at $\Gamma$=0.15. The other parameters are identical to those in Fig.~\ref{fig4}: $J_z = 0.25$, $\beta = 20$ and  $N=6\times6\times3$.
    }
\end{figure}

\section{Loop Length Analysis}
\label{app:loop}

The loop lengths in both the Rydberg chain and Kagome qubit ice are plotted in Fig.~\ref{replyfig2}. Different from the closed-loop algorithm, the loop length becomes larger with the decrease of the transverse field. That's because the number of off-diagonal site operators increases while enlarging the transverse field so that the loop meets the merged operator more frequently and stops earlier. Similarly, when increasing the magnitude of the longitudinal field (detuning) with a fixed transverse field (Rabi frequency), the number of diagonal operators strongly increases, consequently, it would be harder for the loop to stop, and the average length of the loop will increase. 

However, in Fig.~\ref{replyfig2}(b), we can find the decline of the loop when approaching $\Delta=-3$ (corresponding large negative longitudinal field). The possible explanation is that most of the Rydberg atoms stay in the ground state so that the van der Waals interactions have less effect on the diagonal energy. 

In conclusion, different from the loop length in the directed loop algorithm, which is highly relevant to the type of phase (due to its connection to the off-diagonal correlation function), the loop length in our algorithm depends exclusively on the ratio between merged and diagonal operators. That may be the key reason our algorithm shows advantages, especially for a large longitudinal field parameter region.

\begin{figure}[t]
	\centering
    \includegraphics[width=0.99\linewidth]{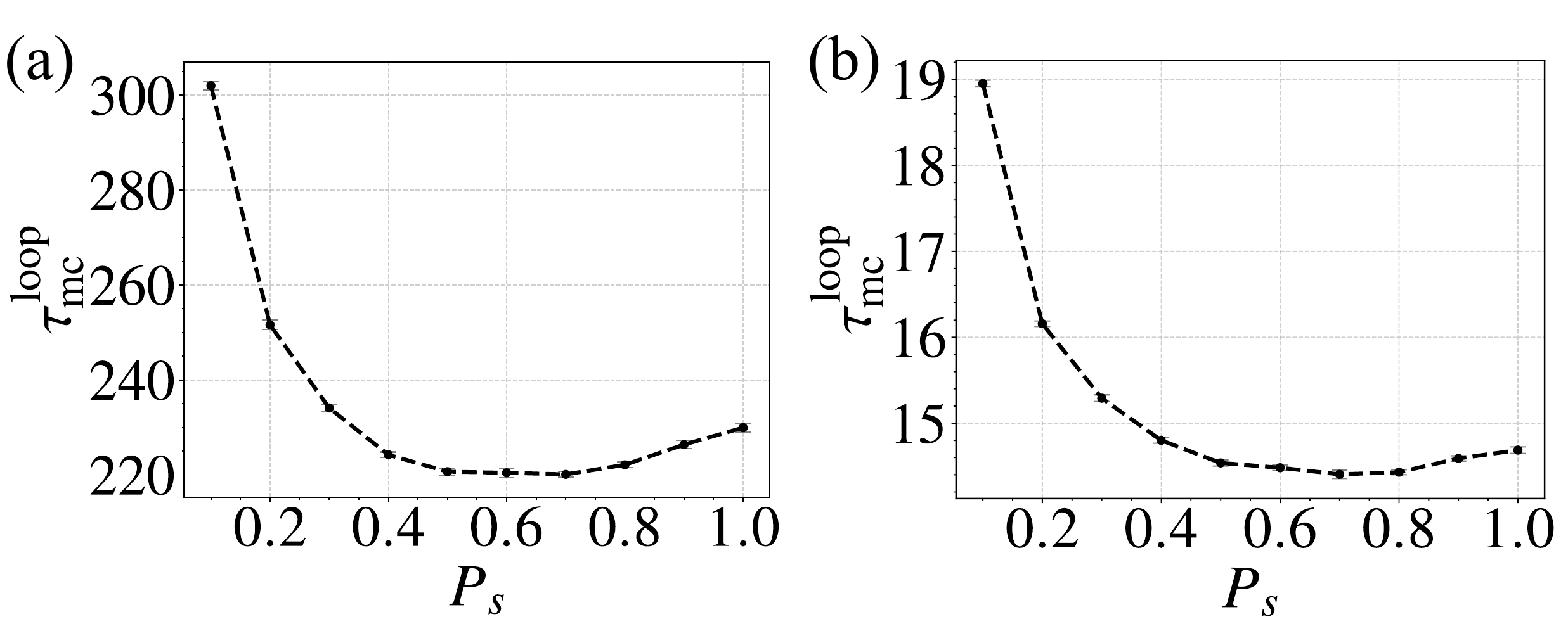}
	\caption{\label{stop_probability} 
    Comparison of the autocorrelation time $\tau^\text{loop}_\text{mc}$ versus the stop probability $P_s$ at two points of Fig.~\ref{fig3}: (a) near the critical point ($\Delta=1$) and (b) deep in the ordered phase ($\Delta=3$). The number of Monte Carlo steps for (b) was increased to $4\times10^7$ to reduce statistical errors.
    }
\end{figure}

\section{Stop Probability}
\label{app:stop_p}
We investigated the relationship between the stop probability, \(P_s\), and the autocorrelation time, \(\tau^\text{loop}_\text{mc}\). The typical behavior of this relationship is shown in Fig.~\ref{stop_probability}. To measure it, we selected two parameter points (\(\Delta=1\) and \(\Delta=3\)) from the phase diagram in Fig.~\ref{fig3}. All other simulation parameters were identical to those used in Fig.~\ref{stop_probability}, except that the number of Monte Carlo steps for the data in panel (b) was increased to \(4\times10^7\) to reduce the statistical error. The parameters corresponding to panel (a) are near the phase transition point, while those for panel (b) lie within the antiferromagnetic phase.

As shown in Fig.~\ref{stop_probability}, \(\tau^\text{loop}_\text{mc}\) initially decreases as \(P_s\) increases, reaches a minimum around \(P_s \approx 0.4\), and then increases again for \(P_s \gtrsim 0.8\). The unbiased stop probability of \(P_s = 0.5\), chosen for our simulations, sits within the region of minimal autocorrelation time.

Our analysis confirms a non-monotonic dependence of the autocorrelation time on the stop probability $P_s$, with an optimum near $P_s \approx 0.6$. As this optimum is system-dependent, we recommend testing a narrow range of $P_s$ values for performance-critical applications.

\bibliography{ref}

\begin{thebibliography}{56}%
\makeatletter
\providecommand \@ifxundefined [1]{%
 \@ifx{#1\undefined}
}%
\providecommand \@ifnum [1]{%
 \ifnum #1\expandafter \@firstoftwo
 \else \expandafter \@secondoftwo
 \fi
}%
\providecommand \@ifx [1]{%
 \ifx #1\expandafter \@firstoftwo
 \else \expandafter \@secondoftwo
 \fi
}%
\providecommand \natexlab [1]{#1}%
\providecommand \enquote  [1]{``#1''}%
\providecommand \bibnamefont  [1]{#1}%
\providecommand \bibfnamefont [1]{#1}%
\providecommand \citenamefont [1]{#1}%
\providecommand \href@noop [0]{\@secondoftwo}%
\providecommand \href [0]{\begingroup \@sanitize@url \@href}%
\providecommand \@href[1]{\@@startlink{#1}\@@href}%
\providecommand \@@href[1]{\endgroup#1\@@endlink}%
\providecommand \@sanitize@url [0]{\catcode `\\12\catcode `\$12\catcode
  `\&12\catcode `\#12\catcode `\^12\catcode `\_12\catcode `\%12\relax}%
\providecommand \@@startlink[1]{}%
\providecommand \@@endlink[0]{}%
\providecommand \url  [0]{\begingroup\@sanitize@url \@url }%
\providecommand \@url [1]{\endgroup\@href {#1}{\urlprefix }}%
\providecommand \urlprefix  [0]{URL }%
\providecommand \Eprint [0]{\href }%
\providecommand \doibase [0]{http://dx.doi.org/}%
\providecommand \selectlanguage [0]{\@gobble}%
\providecommand \bibinfo  [0]{\@secondoftwo}%
\providecommand \bibfield  [0]{\@secondoftwo}%
\providecommand \translation [1]{[#1]}%
\providecommand \BibitemOpen [0]{}%
\providecommand \bibitemStop [0]{}%
\providecommand \bibitemNoStop [0]{.\EOS\space}%
\providecommand \EOS [0]{\spacefactor3000\relax}%
\providecommand \BibitemShut  [1]{\csname bibitem#1\endcsname}%
\let\auto@bib@innerbib\@empty
\bibitem [{\citenamefont {{Pfeuty}}(1970)}]{QTIM_1d}%
  \BibitemOpen
  \bibfield  {author} {\bibinfo {author} {\bibfnamefont {Pierre}\ \bibnamefont
  {{Pfeuty}}},\ }\bibfield  {title} {\enquote {\bibinfo {title} {{The
  one-dimensional Ising model with a transverse field}},}\ }\href {\doibase
  10.1016/0003-4916(70)90270-8} {\bibfield  {journal} {\bibinfo  {journal}
  {Annals of Physics}\ }\textbf {\bibinfo {volume} {57}},\ \bibinfo {pages}
  {79--90} (\bibinfo {year} {1970})}\BibitemShut {NoStop}%
\bibitem [{\citenamefont {Isakov}\ and\ \citenamefont
  {Moessner}(2003)}]{frus_ising_1}%
  \BibitemOpen
  \bibfield  {author} {\bibinfo {author} {\bibfnamefont {S.~V.}\ \bibnamefont
  {Isakov}}\ and\ \bibinfo {author} {\bibfnamefont {R.}~\bibnamefont
  {Moessner}},\ }\bibfield  {title} {\enquote {\bibinfo {title} {Interplay of
  quantum and thermal fluctuations in a frustrated magnet},}\ }\href {\doibase
  10.1103/PhysRevB.68.104409} {\bibfield  {journal} {\bibinfo  {journal} {Phys.
  Rev. B}\ }\textbf {\bibinfo {volume} {68}},\ \bibinfo {pages} {104409}
  (\bibinfo {year} {2003})}\BibitemShut {NoStop}%
\bibitem [{\citenamefont {Moessner}\ \emph
  {et~al.}(2000{\natexlab{a}})\citenamefont {Moessner}, \citenamefont
  {Sondhi},\ and\ \citenamefont {Chandra}}]{frus_ising_2}%
  \BibitemOpen
  \bibfield  {author} {\bibinfo {author} {\bibfnamefont {R.}~\bibnamefont
  {Moessner}}, \bibinfo {author} {\bibfnamefont {S.~L.}\ \bibnamefont
  {Sondhi}}, \ and\ \bibinfo {author} {\bibfnamefont {P.}~\bibnamefont
  {Chandra}},\ }\bibfield  {title} {\enquote {\bibinfo {title} {Two-dimensional
  periodic frustrated ising models in a transverse field},}\ }\href {\doibase
  10.1103/PhysRevLett.84.4457} {\bibfield  {journal} {\bibinfo  {journal}
  {Phys. Rev. Lett.}\ }\textbf {\bibinfo {volume} {84}},\ \bibinfo {pages}
  {4457--4460} (\bibinfo {year} {2000}{\natexlab{a}})}\BibitemShut {NoStop}%
\bibitem [{\citenamefont {{Semeghini}}\ \emph {et~al.}(2021)\citenamefont
  {{Semeghini}}, \citenamefont {{Levine}}, \citenamefont {{Keesling}},
  \citenamefont {{Ebadi}}, \citenamefont {{Wang}}, \citenamefont {{Bluvstein}},
  \citenamefont {{Verresen}}, \citenamefont {{Pichler}}, \citenamefont
  {{Kalinowski}}, \citenamefont {{Samajdar}}, \citenamefont {{Omran}},
  \citenamefont {{Sachdev}}, \citenamefont {{Vishwanath}}, \citenamefont
  {{Greiner}}, \citenamefont {{Vuleti{\'c}}},\ and\ \citenamefont
  {{Lukin}}}]{Rydberg_longr}%
  \BibitemOpen
  \bibfield  {author} {\bibinfo {author} {\bibfnamefont {G.}~\bibnamefont
  {{Semeghini}}}, \bibinfo {author} {\bibfnamefont {H.}~\bibnamefont
  {{Levine}}}, \bibinfo {author} {\bibfnamefont {A.}~\bibnamefont
  {{Keesling}}}, \bibinfo {author} {\bibfnamefont {S.}~\bibnamefont {{Ebadi}}},
  \bibinfo {author} {\bibfnamefont {T.~T.}\ \bibnamefont {{Wang}}}, \bibinfo
  {author} {\bibfnamefont {D.}~\bibnamefont {{Bluvstein}}}, \bibinfo {author}
  {\bibfnamefont {R.}~\bibnamefont {{Verresen}}}, \bibinfo {author}
  {\bibfnamefont {H.}~\bibnamefont {{Pichler}}}, \bibinfo {author}
  {\bibfnamefont {M.}~\bibnamefont {{Kalinowski}}}, \bibinfo {author}
  {\bibfnamefont {R.}~\bibnamefont {{Samajdar}}}, \bibinfo {author}
  {\bibfnamefont {A.}~\bibnamefont {{Omran}}}, \bibinfo {author} {\bibfnamefont
  {S.}~\bibnamefont {{Sachdev}}}, \bibinfo {author} {\bibfnamefont
  {A.}~\bibnamefont {{Vishwanath}}}, \bibinfo {author} {\bibfnamefont
  {M.}~\bibnamefont {{Greiner}}}, \bibinfo {author} {\bibfnamefont
  {V.}~\bibnamefont {{Vuleti{\'c}}}}, \ and\ \bibinfo {author} {\bibfnamefont
  {M.~D.}\ \bibnamefont {{Lukin}}},\ }\bibfield  {title} {\enquote {\bibinfo
  {title} {{Probing topological spin liquids on a programmable quantum
  simulator}},}\ }\href {\doibase 10.1126/science.abi8794} {\bibfield
  {journal} {\bibinfo  {journal} {Science}\ }\textbf {\bibinfo {volume}
  {374}},\ \bibinfo {pages} {1242--1247} (\bibinfo {year} {2021})},\ \Eprint
  {http://arxiv.org/abs/2104.04119} {arXiv:2104.04119 [quant-ph]} \BibitemShut
  {NoStop}%
\bibitem [{\citenamefont {{Zhou}}\ \emph {et~al.}(2021)\citenamefont {{Zhou}},
  \citenamefont {{Zhang}}, \citenamefont {{Pollmann}},\ and\ \citenamefont
  {{You}}}]{fracton01}%
  \BibitemOpen
  \bibfield  {author} {\bibinfo {author} {\bibfnamefont {Zheng}\ \bibnamefont
  {{Zhou}}}, \bibinfo {author} {\bibfnamefont {Xue-Feng}\ \bibnamefont
  {{Zhang}}}, \bibinfo {author} {\bibfnamefont {Frank}\ \bibnamefont
  {{Pollmann}}}, \ and\ \bibinfo {author} {\bibfnamefont {Yizhi}\ \bibnamefont
  {{You}}},\ }\bibfield  {title} {\enquote {\bibinfo {title} {{Fractal Quantum
  Phase Transitions: Critical Phenomena Beyond Renormalization}},}\ }\href
  {\doibase 10.48550/arXiv.2105.05851} {\bibfield  {journal} {\bibinfo
  {journal} {arXiv e-prints}\ ,\ \bibinfo {eid} {arXiv:2105.05851}} (\bibinfo
  {year} {2021})},\ \Eprint {http://arxiv.org/abs/2105.05851} {arXiv:2105.05851
  [cond-mat.str-el]} \BibitemShut {NoStop}%
\bibitem [{\citenamefont {{Zhou}}\ \emph {et~al.}(2023)\citenamefont {{Zhou}},
  \citenamefont {{Liu }}, \citenamefont {{Liu}}, \citenamefont {{Yan}},
  \citenamefont {{Chen }},\ and\ \citenamefont {{Zhang}}}]{frus_ising_3}%
  \BibitemOpen
  \bibfield  {author} {\bibinfo {author} {\bibfnamefont {Zheng}\ \bibnamefont
  {{Zhou}}}, \bibinfo {author} {\bibfnamefont {Changle}\ \bibnamefont {{Liu
  }}}, \bibinfo {author} {\bibfnamefont {Dong-Xu}\ \bibnamefont {{Liu}}},
  \bibinfo {author} {\bibfnamefont {Zheng}\ \bibnamefont {{Yan}}}, \bibinfo
  {author} {\bibfnamefont {Yan}\ \bibnamefont {{Chen }}}, \ and\ \bibinfo
  {author} {\bibfnamefont {Xue-Feng}\ \bibnamefont {{Zhang}}},\ }\bibfield
  {title} {\enquote {\bibinfo {title} {{Quantum tricriticality of
  incommensurate phase induced by quantum strings in frustrated Ising
  magnetism}},}\ }\href {\doibase 10.21468/SciPostPhys.14.3.037} {\bibfield
  {journal} {\bibinfo  {journal} {SciPost Physics}\ }\textbf {\bibinfo {volume}
  {14}},\ \bibinfo {eid} {037} (\bibinfo {year} {2023})},\ \Eprint
  {http://arxiv.org/abs/2005.11133} {arXiv:2005.11133 [cond-mat.str-el]}
  \BibitemShut {NoStop}%
\bibitem [{\citenamefont {Kogut}(1979)}]{lgt_review}%
  \BibitemOpen
  \bibfield  {author} {\bibinfo {author} {\bibfnamefont {John~B.}\ \bibnamefont
  {Kogut}},\ }\bibfield  {title} {\enquote {\bibinfo {title} {An introduction
  to lattice gauge theory and spin systems},}\ }\href {\doibase
  10.1103/RevModPhys.51.659} {\bibfield  {journal} {\bibinfo  {journal} {Rev.
  Mod. Phys.}\ }\textbf {\bibinfo {volume} {51}},\ \bibinfo {pages} {659--713}
  (\bibinfo {year} {1979})}\BibitemShut {NoStop}%
\bibitem [{\citenamefont {Zhou}\ \emph {et~al.}(2025)\citenamefont {Zhou},
  \citenamefont {Yan}, \citenamefont {Liu}, \citenamefont {Chen},\ and\
  \citenamefont {Zhang}}]{lgt_changle}%
  \BibitemOpen
  \bibfield  {author} {\bibinfo {author} {\bibfnamefont {Zheng}\ \bibnamefont
  {Zhou}}, \bibinfo {author} {\bibfnamefont {Zheng}\ \bibnamefont {Yan}},
  \bibinfo {author} {\bibfnamefont {Changle}\ \bibnamefont {Liu}}, \bibinfo
  {author} {\bibfnamefont {Yan}\ \bibnamefont {Chen}}, \ and\ \bibinfo {author}
  {\bibfnamefont {Xue-Feng}\ \bibnamefont {Zhang}},\ }\bibfield  {title}
  {\enquote {\bibinfo {title} {Quantum simulation of two-dimensional u(1) gauge
  theory in rydberg and rydberg-dressed atom arrays},}\ }\href {\doibase
  10.1088/0256-307X/42/5/053705} {\bibfield  {journal} {\bibinfo  {journal}
  {Chinese Physics Letters}\ }\textbf {\bibinfo {volume} {42}},\ \bibinfo
  {pages} {053705} (\bibinfo {year} {2025})}\BibitemShut {NoStop}%
\bibitem [{\citenamefont {{Samajdar}}\ \emph {et~al.}(2023)\citenamefont
  {{Samajdar}}, \citenamefont {{Joshi}}, \citenamefont {{Teng}},\ and\
  \citenamefont {{Sachdev}}}]{Rydberg_gauge}%
  \BibitemOpen
  \bibfield  {author} {\bibinfo {author} {\bibfnamefont {Rhine}\ \bibnamefont
  {{Samajdar}}}, \bibinfo {author} {\bibfnamefont {Darshan~G.}\ \bibnamefont
  {{Joshi}}}, \bibinfo {author} {\bibfnamefont {Yanting}\ \bibnamefont
  {{Teng}}}, \ and\ \bibinfo {author} {\bibfnamefont {Subir}\ \bibnamefont
  {{Sachdev}}},\ }\bibfield  {title} {\enquote {\bibinfo {title} {{Emergent
  Z$_{2}$ Gauge Theories and Topological Excitations in Rydberg Atom
  Arrays}},}\ }\href {\doibase 10.1103/PhysRevLett.130.043601} {\bibfield
  {journal} {\bibinfo  {journal} {\prl}\ }\textbf {\bibinfo {volume} {130}},\
  \bibinfo {eid} {043601} (\bibinfo {year} {2023})},\ \Eprint
  {http://arxiv.org/abs/2204.00632} {arXiv:2204.00632 [cond-mat.quant-gas]}
  \BibitemShut {NoStop}%
\bibitem [{\citenamefont {{Yan}}\ \emph {et~al.}(2023)\citenamefont {{Yan}},
  \citenamefont {{Wang}}, \citenamefont {{Samajdar}}, \citenamefont
  {{Sachdev}},\ and\ \citenamefont {{Meng}}}]{Rydberg_glass}%
  \BibitemOpen
  \bibfield  {author} {\bibinfo {author} {\bibfnamefont {Zheng}\ \bibnamefont
  {{Yan}}}, \bibinfo {author} {\bibfnamefont {Yan-Cheng}\ \bibnamefont
  {{Wang}}}, \bibinfo {author} {\bibfnamefont {Rhine}\ \bibnamefont
  {{Samajdar}}}, \bibinfo {author} {\bibfnamefont {Subir}\ \bibnamefont
  {{Sachdev}}}, \ and\ \bibinfo {author} {\bibfnamefont {Zi~Yang}\ \bibnamefont
  {{Meng}}},\ }\bibfield  {title} {\enquote {\bibinfo {title} {{Emergent Glassy
  Behavior in a Kagome Rydberg Atom Array}},}\ }\href {\doibase
  10.1103/PhysRevLett.130.206501} {\bibfield  {journal} {\bibinfo  {journal}
  {\prl}\ }\textbf {\bibinfo {volume} {130}},\ \bibinfo {eid} {206501}
  (\bibinfo {year} {2023})},\ \Eprint {http://arxiv.org/abs/2301.07127}
  {arXiv:2301.07127 [cond-mat.quant-gas]} \BibitemShut {NoStop}%
\bibitem [{\citenamefont {{Wiedmann}}\ \emph {et~al.}(2024)\citenamefont
  {{Wiedmann}}, \citenamefont {{Lenke}}, \citenamefont {{M{\"u}hlhauser}},\
  and\ \citenamefont {{Schmidt}}}]{fracton02}%
  \BibitemOpen
  \bibfield  {author} {\bibinfo {author} {\bibfnamefont {Raymond}\ \bibnamefont
  {{Wiedmann}}}, \bibinfo {author} {\bibfnamefont {Lea}\ \bibnamefont
  {{Lenke}}}, \bibinfo {author} {\bibfnamefont {Matthias}\ \bibnamefont
  {{M{\"u}hlhauser}}}, \ and\ \bibinfo {author} {\bibfnamefont {Kai~Phillip}\
  \bibnamefont {{Schmidt}}},\ }\bibfield  {title} {\enquote {\bibinfo {title}
  {{Absence of fractal quantum criticality in the quantum Newman-Moore
  model}},}\ }\href {\doibase 10.1103/PhysRevResearch.6.013191} {\bibfield
  {journal} {\bibinfo  {journal} {Physical Review Research}\ }\textbf {\bibinfo
  {volume} {6}},\ \bibinfo {eid} {013191} (\bibinfo {year} {2024})},\ \Eprint
  {http://arxiv.org/abs/2302.01773} {arXiv:2302.01773 [cond-mat.str-el]}
  \BibitemShut {NoStop}%
\bibitem [{\citenamefont {Baxter}(1982)}]{Baxter}%
  \BibitemOpen
  \bibfield  {author} {\bibinfo {author} {\bibfnamefont {R.~J.}\ \bibnamefont
  {Baxter}},\ }\href {\doibase 10.1142/9789814415255_0002} {\emph {\bibinfo
  {title} {{Exactly solved models in statistical mechanics}}}}\ (\bibinfo
  {publisher} {Dover Publications, INC. Mineola, New York},\ \bibinfo {year}
  {1982})\BibitemShut {NoStop}%
\bibitem [{\citenamefont {Li}\ \emph {et~al.}(2020{\natexlab{a}})\citenamefont
  {Li}, \citenamefont {Da~Liao}, \citenamefont {Chen}, \citenamefont {Zeng},
  \citenamefont {Sheng}, \citenamefont {Qi}, \citenamefont {Meng},\ and\
  \citenamefont {Li}}]{tmgo_kt}%
  \BibitemOpen
  \bibfield  {author} {\bibinfo {author} {\bibfnamefont {Han}\ \bibnamefont
  {Li}}, \bibinfo {author} {\bibfnamefont {Yuan}\ \bibnamefont {Da~Liao}},
  \bibinfo {author} {\bibfnamefont {Bin-Bin}\ \bibnamefont {Chen}}, \bibinfo
  {author} {\bibfnamefont {Xu-Tao}\ \bibnamefont {Zeng}}, \bibinfo {author}
  {\bibfnamefont {Xian-Lei}\ \bibnamefont {Sheng}}, \bibinfo {author}
  {\bibfnamefont {Yang}\ \bibnamefont {Qi}}, \bibinfo {author} {\bibfnamefont
  {Zi~Yang}\ \bibnamefont {Meng}}, \ and\ \bibinfo {author} {\bibfnamefont
  {Wei}\ \bibnamefont {Li}},\ }\bibfield  {title} {\enquote {\bibinfo {title}
  {{K}osterlitz-{T}houless melting of magnetic order in the triangular quantum
  {I}sing material {T}m{M}g{G}a{O}$_4$},}\ }\href {\doibase
  10.1038/s41467-020-14907-8} {\bibfield  {journal} {\bibinfo  {journal} {Nat.
  Commun.}\ }\textbf {\bibinfo {volume} {11}},\ \bibinfo {pages} {1--8}
  (\bibinfo {year} {2020}{\natexlab{a}})}\BibitemShut {NoStop}%
\bibitem [{\citenamefont {Liu}\ \emph {et~al.}(2020{\natexlab{a}})\citenamefont
  {Liu}, \citenamefont {Huang},\ and\ \citenamefont {Chen}}]{tmgo_model}%
  \BibitemOpen
  \bibfield  {author} {\bibinfo {author} {\bibfnamefont {Changle}\ \bibnamefont
  {Liu}}, \bibinfo {author} {\bibfnamefont {Chun-Jiong}\ \bibnamefont {Huang}},
  \ and\ \bibinfo {author} {\bibfnamefont {Gang}\ \bibnamefont {Chen}},\
  }\bibfield  {title} {\enquote {\bibinfo {title} {Intrinsic quantum {I}sing
  model on a triangular lattice magnet
  $\mathrm{Tm}\mathrm{Mg}\mathrm{Ga}\mathrm{O}_{4}$},}\ }\href {\doibase
  10.1103/PhysRevResearch.2.043013} {\bibfield  {journal} {\bibinfo  {journal}
  {Phys. Rev. Research}\ }\textbf {\bibinfo {volume} {2}},\ \bibinfo {pages}
  {043013} (\bibinfo {year} {2020}{\natexlab{a}})}\BibitemShut {NoStop}%
\bibitem [{\citenamefont {Shen}\ \emph {et~al.}(2019)\citenamefont {Shen},
  \citenamefont {Liu}, \citenamefont {Qin}, \citenamefont {Shen}, \citenamefont
  {Li}, \citenamefont {Bewley}, \citenamefont {Schneidewind}, \citenamefont
  {Chen},\ and\ \citenamefont {Zhao}}]{tmgo_neu}%
  \BibitemOpen
  \bibfield  {author} {\bibinfo {author} {\bibfnamefont {Yao}\ \bibnamefont
  {Shen}}, \bibinfo {author} {\bibfnamefont {Changle}\ \bibnamefont {Liu}},
  \bibinfo {author} {\bibfnamefont {Yayuan}\ \bibnamefont {Qin}}, \bibinfo
  {author} {\bibfnamefont {Shoudong}\ \bibnamefont {Shen}}, \bibinfo {author}
  {\bibfnamefont {Yao-Dong}\ \bibnamefont {Li}}, \bibinfo {author}
  {\bibfnamefont {Robert}\ \bibnamefont {Bewley}}, \bibinfo {author}
  {\bibfnamefont {Astrid}\ \bibnamefont {Schneidewind}}, \bibinfo {author}
  {\bibfnamefont {Gang}\ \bibnamefont {Chen}}, \ and\ \bibinfo {author}
  {\bibfnamefont {Jun}\ \bibnamefont {Zhao}},\ }\bibfield  {title} {\enquote
  {\bibinfo {title} {Intertwined dipolar and multipolar order in the
  triangular-lattice magnet {T}m{M}g{G}a{O}$_4$},}\ }\href {\doibase
  10.1038/s41467-019-12410-3} {\bibfield  {journal} {\bibinfo  {journal} {Nat.
  Commun.}\ }\textbf {\bibinfo {volume} {10}},\ \bibinfo {pages} {1--7}
  (\bibinfo {year} {2019})}\BibitemShut {NoStop}%
\bibitem [{\citenamefont {Li}\ \emph {et~al.}(2020{\natexlab{b}})\citenamefont
  {Li}, \citenamefont {Bachus}, \citenamefont {Deng}, \citenamefont {Schmidt},
  \citenamefont {Thoma}, \citenamefont {Hutanu}, \citenamefont {Tokiwa},
  \citenamefont {Tsirlin},\ and\ \citenamefont {Gegenwart}}]{tmgo_uud}%
  \BibitemOpen
  \bibfield  {author} {\bibinfo {author} {\bibfnamefont {Yuesheng}\
  \bibnamefont {Li}}, \bibinfo {author} {\bibfnamefont {Sebastian}\
  \bibnamefont {Bachus}}, \bibinfo {author} {\bibfnamefont {Hao}\ \bibnamefont
  {Deng}}, \bibinfo {author} {\bibfnamefont {Wolfgang}\ \bibnamefont
  {Schmidt}}, \bibinfo {author} {\bibfnamefont {Henrik}\ \bibnamefont {Thoma}},
  \bibinfo {author} {\bibfnamefont {Vladimir}\ \bibnamefont {Hutanu}}, \bibinfo
  {author} {\bibfnamefont {Yoshifumi}\ \bibnamefont {Tokiwa}}, \bibinfo
  {author} {\bibfnamefont {Alexander~A.}\ \bibnamefont {Tsirlin}}, \ and\
  \bibinfo {author} {\bibfnamefont {Philipp}\ \bibnamefont {Gegenwart}},\
  }\bibfield  {title} {\enquote {\bibinfo {title} {Partial up-up-down order
  with the continuously distributed order parameter in the triangular
  antiferromagnet {T}m{M}g{G}a{O}$_4$},}\ }\href {\doibase
  10.1103/PhysRevX.10.011007} {\bibfield  {journal} {\bibinfo  {journal} {Phys.
  Rev. X}\ }\textbf {\bibinfo {volume} {10}},\ \bibinfo {pages} {011007}
  (\bibinfo {year} {2020}{\natexlab{b}})}\BibitemShut {NoStop}%
\bibitem [{\citenamefont {{Guo}}\ \emph {et~al.}(2024)\citenamefont {{Guo}},
  \citenamefont {{Wu}}, \citenamefont {{Ye}}, \citenamefont {{Zhang}},
  \citenamefont {{Lian}}, \citenamefont {{Yao}}, \citenamefont {{Wang}},
  \citenamefont {{Yan}}, \citenamefont {{Yi}}, \citenamefont {{Xu}},
  \citenamefont {{Li}}, \citenamefont {{Hou}}, \citenamefont {{Xu}},
  \citenamefont {{Guo}}, \citenamefont {{Zhang}}, \citenamefont {{Qi}},
  \citenamefont {{Zhou}}, \citenamefont {{He}},\ and\ \citenamefont
  {{Duan}}}]{trap_ion}%
  \BibitemOpen
  \bibfield  {author} {\bibinfo {author} {\bibfnamefont {S.~A.}\ \bibnamefont
  {{Guo}}}, \bibinfo {author} {\bibfnamefont {Y.~K.}\ \bibnamefont {{Wu}}},
  \bibinfo {author} {\bibfnamefont {J.}~\bibnamefont {{Ye}}}, \bibinfo {author}
  {\bibfnamefont {L.}~\bibnamefont {{Zhang}}}, \bibinfo {author} {\bibfnamefont
  {W.~Q.}\ \bibnamefont {{Lian}}}, \bibinfo {author} {\bibfnamefont
  {R.}~\bibnamefont {{Yao}}}, \bibinfo {author} {\bibfnamefont
  {Y.}~\bibnamefont {{Wang}}}, \bibinfo {author} {\bibfnamefont {R.~Y.}\
  \bibnamefont {{Yan}}}, \bibinfo {author} {\bibfnamefont {Y.~J.}\ \bibnamefont
  {{Yi}}}, \bibinfo {author} {\bibfnamefont {Y.~L.}\ \bibnamefont {{Xu}}},
  \bibinfo {author} {\bibfnamefont {B.~W.}\ \bibnamefont {{Li}}}, \bibinfo
  {author} {\bibfnamefont {Y.~H.}\ \bibnamefont {{Hou}}}, \bibinfo {author}
  {\bibfnamefont {Y.~Z.}\ \bibnamefont {{Xu}}}, \bibinfo {author}
  {\bibfnamefont {W.~X.}\ \bibnamefont {{Guo}}}, \bibinfo {author}
  {\bibfnamefont {C.}~\bibnamefont {{Zhang}}}, \bibinfo {author} {\bibfnamefont
  {B.~X.}\ \bibnamefont {{Qi}}}, \bibinfo {author} {\bibfnamefont {Z.~C.}\
  \bibnamefont {{Zhou}}}, \bibinfo {author} {\bibfnamefont {L.}~\bibnamefont
  {{He}}}, \ and\ \bibinfo {author} {\bibfnamefont {L.~M.}\ \bibnamefont
  {{Duan}}},\ }\bibfield  {title} {\enquote {\bibinfo {title} {{A site-resolved
  two-dimensional quantum simulator with hundreds of trapped ions}},}\ }\href
  {\doibase 10.1038/s41586-024-07459-0} {\bibfield  {journal} {\bibinfo
  {journal} {\nat}\ }\textbf {\bibinfo {volume} {630}},\ \bibinfo {pages}
  {613--618} (\bibinfo {year} {2024})},\ \Eprint
  {http://arxiv.org/abs/2311.17163} {arXiv:2311.17163 [quant-ph]} \BibitemShut
  {NoStop}%
\bibitem [{\citenamefont {Shen}\ \emph {et~al.}(2016)\citenamefont {Shen},
  \citenamefont {Wu}, \citenamefont {Song}, \citenamefont {Sun}, \citenamefont
  {Yang}, \citenamefont {Chai}, \citenamefont {Shang}, \citenamefont {Wang},
  \citenamefont {Scott},\ and\ \citenamefont {Sun}}]{chai}%
  \BibitemOpen
  \bibfield  {author} {\bibinfo {author} {\bibfnamefont {Shi-Peng}\
  \bibnamefont {Shen}}, \bibinfo {author} {\bibfnamefont {Jia-Chuan}\
  \bibnamefont {Wu}}, \bibinfo {author} {\bibfnamefont {Jun-Da}\ \bibnamefont
  {Song}}, \bibinfo {author} {\bibfnamefont {Xue-Feng}\ \bibnamefont {Sun}},
  \bibinfo {author} {\bibfnamefont {Yi-Feng}\ \bibnamefont {Yang}}, \bibinfo
  {author} {\bibfnamefont {Yi-Sheng}\ \bibnamefont {Chai}}, \bibinfo {author}
  {\bibfnamefont {Da-Shan}\ \bibnamefont {Shang}}, \bibinfo {author}
  {\bibfnamefont {Shou-Guo}\ \bibnamefont {Wang}}, \bibinfo {author}
  {\bibfnamefont {James~F}\ \bibnamefont {Scott}}, \ and\ \bibinfo {author}
  {\bibfnamefont {Young}\ \bibnamefont {Sun}},\ }\bibfield  {title} {\enquote
  {\bibinfo {title} {{Quantum electric-dipole liquid on a triangular
  lattice}},}\ }\href {\doibase 10.1038/ncomms10569} {\bibfield  {journal}
  {\bibinfo  {journal} {Nat. Commun.}\ }\textbf {\bibinfo {volume} {7}},\
  \bibinfo {pages} {10569} (\bibinfo {year} {2016})}\BibitemShut {NoStop}%
\bibitem [{\citenamefont {{Bernien}}\ \emph {et~al.}(2017)\citenamefont
  {{Bernien}}, \citenamefont {{Schwartz}}, \citenamefont {{Keesling}},
  \citenamefont {{Levine}}, \citenamefont {{Omran}}, \citenamefont {{Pichler}},
  \citenamefont {{Choi}}, \citenamefont {{Zibrov}}, \citenamefont {{Endres}},
  \citenamefont {{Greiner}}, \citenamefont {{Vuleti{\'c}}},\ and\ \citenamefont
  {{Lukin}}}]{Rydberg_chain}%
  \BibitemOpen
  \bibfield  {author} {\bibinfo {author} {\bibfnamefont {Hannes}\ \bibnamefont
  {{Bernien}}}, \bibinfo {author} {\bibfnamefont {Sylvain}\ \bibnamefont
  {{Schwartz}}}, \bibinfo {author} {\bibfnamefont {Alexander}\ \bibnamefont
  {{Keesling}}}, \bibinfo {author} {\bibfnamefont {Harry}\ \bibnamefont
  {{Levine}}}, \bibinfo {author} {\bibfnamefont {Ahmed}\ \bibnamefont
  {{Omran}}}, \bibinfo {author} {\bibfnamefont {Hannes}\ \bibnamefont
  {{Pichler}}}, \bibinfo {author} {\bibfnamefont {Soonwon}\ \bibnamefont
  {{Choi}}}, \bibinfo {author} {\bibfnamefont {Alexander~S.}\ \bibnamefont
  {{Zibrov}}}, \bibinfo {author} {\bibfnamefont {Manuel}\ \bibnamefont
  {{Endres}}}, \bibinfo {author} {\bibfnamefont {Markus}\ \bibnamefont
  {{Greiner}}}, \bibinfo {author} {\bibfnamefont {Vladan}\ \bibnamefont
  {{Vuleti{\'c}}}}, \ and\ \bibinfo {author} {\bibfnamefont {Mikhail~D.}\
  \bibnamefont {{Lukin}}},\ }\bibfield  {title} {\enquote {\bibinfo {title}
  {{Probing many-body dynamics on a 51-atom quantum simulator}},}\ }\href
  {\doibase 10.1038/nature24622} {\bibfield  {journal} {\bibinfo  {journal}
  {\nat}\ }\textbf {\bibinfo {volume} {551}},\ \bibinfo {pages} {579--584}
  (\bibinfo {year} {2017})},\ \Eprint {http://arxiv.org/abs/1707.04344}
  {arXiv:1707.04344 [quant-ph]} \BibitemShut {NoStop}%
\bibitem [{\citenamefont {{Ebadi}}\ \emph {et~al.}(2021)\citenamefont
  {{Ebadi}}, \citenamefont {{Wang}}, \citenamefont {{Levine}}, \citenamefont
  {{Keesling}}, \citenamefont {{Semeghini}}, \citenamefont {{Omran}},
  \citenamefont {{Bluvstein}}, \citenamefont {{Samajdar}}, \citenamefont
  {{Pichler}}, \citenamefont {{Ho}}, \citenamefont {{Choi}}, \citenamefont
  {{Sachdev}}, \citenamefont {{Greiner}}, \citenamefont {{Vuleti{\'c}}},\ and\
  \citenamefont {{Lukin}}}]{Rydberg_nature1}%
  \BibitemOpen
  \bibfield  {author} {\bibinfo {author} {\bibfnamefont {Sepehr}\ \bibnamefont
  {{Ebadi}}}, \bibinfo {author} {\bibfnamefont {Tout~T.}\ \bibnamefont
  {{Wang}}}, \bibinfo {author} {\bibfnamefont {Harry}\ \bibnamefont
  {{Levine}}}, \bibinfo {author} {\bibfnamefont {Alexander}\ \bibnamefont
  {{Keesling}}}, \bibinfo {author} {\bibfnamefont {Giulia}\ \bibnamefont
  {{Semeghini}}}, \bibinfo {author} {\bibfnamefont {Ahmed}\ \bibnamefont
  {{Omran}}}, \bibinfo {author} {\bibfnamefont {Dolev}\ \bibnamefont
  {{Bluvstein}}}, \bibinfo {author} {\bibfnamefont {Rhine}\ \bibnamefont
  {{Samajdar}}}, \bibinfo {author} {\bibfnamefont {Hannes}\ \bibnamefont
  {{Pichler}}}, \bibinfo {author} {\bibfnamefont {Wen~Wei}\ \bibnamefont
  {{Ho}}}, \bibinfo {author} {\bibfnamefont {Soonwon}\ \bibnamefont {{Choi}}},
  \bibinfo {author} {\bibfnamefont {Subir}\ \bibnamefont {{Sachdev}}}, \bibinfo
  {author} {\bibfnamefont {Markus}\ \bibnamefont {{Greiner}}}, \bibinfo
  {author} {\bibfnamefont {Vladan}\ \bibnamefont {{Vuleti{\'c}}}}, \ and\
  \bibinfo {author} {\bibfnamefont {Mikhail~D.}\ \bibnamefont {{Lukin}}},\
  }\bibfield  {title} {\enquote {\bibinfo {title} {{Quantum phases of matter on
  a 256-atom programmable quantum simulator}},}\ }\href {\doibase
  10.1038/s41586-021-03582-4} {\bibfield  {journal} {\bibinfo  {journal}
  {\nat}\ }\textbf {\bibinfo {volume} {595}},\ \bibinfo {pages} {227--232}
  (\bibinfo {year} {2021})},\ \Eprint {http://arxiv.org/abs/2012.12281}
  {arXiv:2012.12281 [quant-ph]} \BibitemShut {NoStop}%
\bibitem [{\citenamefont {{Scholl}}\ \emph {et~al.}(2021)\citenamefont
  {{Scholl}}, \citenamefont {{Schuler}}, \citenamefont {{Williams}},
  \citenamefont {{Eberharter}}, \citenamefont {{Barredo}}, \citenamefont
  {{Schymik}}, \citenamefont {{Lienhard}}, \citenamefont {{Henry}},
  \citenamefont {{Lang}}, \citenamefont {{Lahaye}}, \citenamefont
  {{L{\"a}uchli}},\ and\ \citenamefont {{Browaeys}}}]{Rydberg_nature2}%
  \BibitemOpen
  \bibfield  {author} {\bibinfo {author} {\bibfnamefont {Pascal}\ \bibnamefont
  {{Scholl}}}, \bibinfo {author} {\bibfnamefont {Michael}\ \bibnamefont
  {{Schuler}}}, \bibinfo {author} {\bibfnamefont {Hannah~J.}\ \bibnamefont
  {{Williams}}}, \bibinfo {author} {\bibfnamefont {Alexander~A.}\ \bibnamefont
  {{Eberharter}}}, \bibinfo {author} {\bibfnamefont {Daniel}\ \bibnamefont
  {{Barredo}}}, \bibinfo {author} {\bibfnamefont {Kai-Niklas}\ \bibnamefont
  {{Schymik}}}, \bibinfo {author} {\bibfnamefont {Vincent}\ \bibnamefont
  {{Lienhard}}}, \bibinfo {author} {\bibfnamefont {Louis-Paul}\ \bibnamefont
  {{Henry}}}, \bibinfo {author} {\bibfnamefont {Thomas~C.}\ \bibnamefont
  {{Lang}}}, \bibinfo {author} {\bibfnamefont {Thierry}\ \bibnamefont
  {{Lahaye}}}, \bibinfo {author} {\bibfnamefont {Andreas~M.}\ \bibnamefont
  {{L{\"a}uchli}}}, \ and\ \bibinfo {author} {\bibfnamefont {Antoine}\
  \bibnamefont {{Browaeys}}},\ }\bibfield  {title} {\enquote {\bibinfo {title}
  {{Quantum simulation of 2D antiferromagnets with hundreds of Rydberg
  atoms}},}\ }\href {\doibase 10.1038/s41586-021-03585-1} {\bibfield  {journal}
  {\bibinfo  {journal} {\nat}\ }\textbf {\bibinfo {volume} {595}},\ \bibinfo
  {pages} {233--238} (\bibinfo {year} {2021})},\ \Eprint
  {http://arxiv.org/abs/2012.12268} {arXiv:2012.12268 [quant-ph]} \BibitemShut
  {NoStop}%
\bibitem [{\citenamefont {King}\ \emph {et~al.}(2018)\citenamefont {King},
  \citenamefont {Carrasquilla}, \citenamefont {Raymond}, \citenamefont
  {Ozfidan}, \citenamefont {Andriyash}, \citenamefont {Berkley}, \citenamefont
  {Reis}, \citenamefont {Lanting}, \citenamefont {Harris}, \citenamefont
  {Altomare}, \citenamefont {Boothby}, \citenamefont {Bunyk}, \citenamefont
  {Enderud}, \citenamefont {Fr{\'{e}}chette}, \citenamefont {Hoskinson},
  \citenamefont {Ladizinsky}, \citenamefont {Oh}, \citenamefont
  {Poulin-Lamarre}, \citenamefont {Rich}, \citenamefont {Sato}, \citenamefont
  {Smirnov}, \citenamefont {Swenson}, \citenamefont {Volkmann}, \citenamefont
  {Whittaker}, \citenamefont {Yao}, \citenamefont {Ladizinsky}, \citenamefont
  {Johnson}, \citenamefont {Hilton},\ and\ \citenamefont {Amin}}]{dwave}%
  \BibitemOpen
  \bibfield  {author} {\bibinfo {author} {\bibfnamefont {Andrew~D}\
  \bibnamefont {King}}, \bibinfo {author} {\bibfnamefont {Juan}\ \bibnamefont
  {Carrasquilla}}, \bibinfo {author} {\bibfnamefont {Jack}\ \bibnamefont
  {Raymond}}, \bibinfo {author} {\bibfnamefont {Isil}\ \bibnamefont {Ozfidan}},
  \bibinfo {author} {\bibfnamefont {Evgeny}\ \bibnamefont {Andriyash}},
  \bibinfo {author} {\bibfnamefont {Andrew}\ \bibnamefont {Berkley}}, \bibinfo
  {author} {\bibfnamefont {Mauricio}\ \bibnamefont {Reis}}, \bibinfo {author}
  {\bibfnamefont {Trevor}\ \bibnamefont {Lanting}}, \bibinfo {author}
  {\bibfnamefont {Richard}\ \bibnamefont {Harris}}, \bibinfo {author}
  {\bibfnamefont {Fabio}\ \bibnamefont {Altomare}}, \bibinfo {author}
  {\bibfnamefont {Kelly}\ \bibnamefont {Boothby}}, \bibinfo {author}
  {\bibfnamefont {Paul~I}\ \bibnamefont {Bunyk}}, \bibinfo {author}
  {\bibfnamefont {Colin}\ \bibnamefont {Enderud}}, \bibinfo {author}
  {\bibfnamefont {Alexandre}\ \bibnamefont {Fr{\'{e}}chette}}, \bibinfo
  {author} {\bibfnamefont {Emile}\ \bibnamefont {Hoskinson}}, \bibinfo {author}
  {\bibfnamefont {Nicolas}\ \bibnamefont {Ladizinsky}}, \bibinfo {author}
  {\bibfnamefont {Travis}\ \bibnamefont {Oh}}, \bibinfo {author} {\bibfnamefont
  {Gabriel}\ \bibnamefont {Poulin-Lamarre}}, \bibinfo {author} {\bibfnamefont
  {Christopher}\ \bibnamefont {Rich}}, \bibinfo {author} {\bibfnamefont {Yuki}\
  \bibnamefont {Sato}}, \bibinfo {author} {\bibfnamefont {Anatoly~Yu.}\
  \bibnamefont {Smirnov}}, \bibinfo {author} {\bibfnamefont {Loren~J}\
  \bibnamefont {Swenson}}, \bibinfo {author} {\bibfnamefont {Mark~H}\
  \bibnamefont {Volkmann}}, \bibinfo {author} {\bibfnamefont {Jed}\
  \bibnamefont {Whittaker}}, \bibinfo {author} {\bibfnamefont {Jason}\
  \bibnamefont {Yao}}, \bibinfo {author} {\bibfnamefont {Eric}\ \bibnamefont
  {Ladizinsky}}, \bibinfo {author} {\bibfnamefont {Mark~W}\ \bibnamefont
  {Johnson}}, \bibinfo {author} {\bibfnamefont {Jeremy}\ \bibnamefont
  {Hilton}}, \ and\ \bibinfo {author} {\bibfnamefont {Mohammad~H}\ \bibnamefont
  {Amin}},\ }\bibfield  {title} {\enquote {\bibinfo {title} {{Observation of
  topological phenomena in a programmable lattice of 1,800 qubits}},}\ }\href
  {\doibase 10.1038/s41586-018-0410-x} {\bibfield  {journal} {\bibinfo
  {journal} {Nature}\ }\textbf {\bibinfo {volume} {560}},\ \bibinfo {pages}
  {456--460} (\bibinfo {year} {2018})}\BibitemShut {NoStop}%
\bibitem [{\citenamefont {Narasimhan}\ \emph {et~al.}(2024)\citenamefont
  {Narasimhan}, \citenamefont {Humeniuk}, \citenamefont {Roy},\ and\
  \citenamefont {Drouin-Touchette}}]{Kagome_dwave}%
  \BibitemOpen
  \bibfield  {author} {\bibinfo {author} {\bibfnamefont {Pratyankara}\
  \bibnamefont {Narasimhan}}, \bibinfo {author} {\bibfnamefont {Stephan}\
  \bibnamefont {Humeniuk}}, \bibinfo {author} {\bibfnamefont {Ananda}\
  \bibnamefont {Roy}}, \ and\ \bibinfo {author} {\bibfnamefont {Victor}\
  \bibnamefont {Drouin-Touchette}},\ }\bibfield  {title} {\enquote {\bibinfo
  {title} {Simulating the transverse-field ising model on the kagome lattice
  using a programmable quantum annealer},}\ }\href {\doibase
  10.1103/PhysRevB.110.054432} {\bibfield  {journal} {\bibinfo  {journal}
  {Phys. Rev. B}\ }\textbf {\bibinfo {volume} {110}},\ \bibinfo {pages}
  {054432} (\bibinfo {year} {2024})}\BibitemShut {NoStop}%
\bibitem [{\citenamefont {{Lopez-Bezanilla}}\ \emph {et~al.}(2023)\citenamefont
  {{Lopez-Bezanilla}}, \citenamefont {{Raymond}}, \citenamefont {{Boothby}},
  \citenamefont {{Carrasquilla}}, \citenamefont {{Nisoli}},\ and\ \citenamefont
  {{King}}}]{kagome_dwave2}%
  \BibitemOpen
  \bibfield  {author} {\bibinfo {author} {\bibfnamefont {Alejandro}\
  \bibnamefont {{Lopez-Bezanilla}}}, \bibinfo {author} {\bibfnamefont {Jack}\
  \bibnamefont {{Raymond}}}, \bibinfo {author} {\bibfnamefont {Kelly}\
  \bibnamefont {{Boothby}}}, \bibinfo {author} {\bibfnamefont {Juan}\
  \bibnamefont {{Carrasquilla}}}, \bibinfo {author} {\bibfnamefont {Cristiano}\
  \bibnamefont {{Nisoli}}}, \ and\ \bibinfo {author} {\bibfnamefont
  {Andrew~D.}\ \bibnamefont {{King}}},\ }\bibfield  {title} {\enquote {\bibinfo
  {title} {{Kagome qubit ice}},}\ }\href {\doibase 10.1038/s41467-023-36760-1}
  {\bibfield  {journal} {\bibinfo  {journal} {Nature Communications}\ }\textbf
  {\bibinfo {volume} {14}},\ \bibinfo {eid} {1105} (\bibinfo {year} {2023})},\
  \Eprint {http://arxiv.org/abs/2301.01853} {arXiv:2301.01853
  [cond-mat.stat-mech]} \BibitemShut {NoStop}%
\bibitem [{\citenamefont {Sylju\aa{}sen}\ and\ \citenamefont
  {Sandvik}(2002)}]{SSE_Sandvik_2}%
  \BibitemOpen
  \bibfield  {author} {\bibinfo {author} {\bibfnamefont {Olav~F.}\ \bibnamefont
  {Sylju\aa{}sen}}\ and\ \bibinfo {author} {\bibfnamefont {Anders~W.}\
  \bibnamefont {Sandvik}},\ }\bibfield  {title} {\enquote {\bibinfo {title}
  {Quantum monte carlo with directed loops},}\ }\href {\doibase
  10.1103/PhysRevE.66.046701} {\bibfield  {journal} {\bibinfo  {journal} {Phys.
  Rev. E}\ }\textbf {\bibinfo {volume} {66}},\ \bibinfo {pages} {046701}
  (\bibinfo {year} {2002})}\BibitemShut {NoStop}%
\bibitem [{\citenamefont {Syljuåsen}(2003)}]{Syljuåsen_2003}%
  \BibitemOpen
  \bibfield  {author} {\bibinfo {author} {\bibfnamefont {Olav~F.}\ \bibnamefont
  {Syljuåsen}},\ }\bibfield  {title} {\enquote {\bibinfo {title} {Directed
  loop updates for quantum lattice models},}\ }\href {\doibase
  10.1103/PhysRevE.67.046701} {\bibfield  {journal} {\bibinfo  {journal}
  {Physical Review E}\ }\textbf {\bibinfo {volume} {67}},\ \bibinfo {pages}
  {046701} (\bibinfo {year} {2003})}\BibitemShut {NoStop}%
\bibitem [{\citenamefont {Sandvik}(2003)}]{SandVik_SSEforQIM}%
  \BibitemOpen
  \bibfield  {author} {\bibinfo {author} {\bibfnamefont {Anders~W.}\
  \bibnamefont {Sandvik}},\ }\bibfield  {title} {\enquote {\bibinfo {title}
  {Stochastic series expansion method for quantum ising models with arbitrary
  interactions},}\ }\href {\doibase 10.1103/PhysRevE.68.056701} {\bibfield
  {journal} {\bibinfo  {journal} {Phys. Rev. E}\ }\textbf {\bibinfo {volume}
  {68}},\ \bibinfo {pages} {056701} (\bibinfo {year} {2003})}\BibitemShut
  {NoStop}%
\bibitem [{\citenamefont {Moessner}\ \emph
  {et~al.}(2000{\natexlab{b}})\citenamefont {Moessner}, \citenamefont
  {Sondhi},\ and\ \citenamefont {Chandra}}]{PhysRevLett.84.4457}%
  \BibitemOpen
  \bibfield  {author} {\bibinfo {author} {\bibfnamefont {R.}~\bibnamefont
  {Moessner}}, \bibinfo {author} {\bibfnamefont {S.~L.}\ \bibnamefont
  {Sondhi}}, \ and\ \bibinfo {author} {\bibfnamefont {P.}~\bibnamefont
  {Chandra}},\ }\bibfield  {title} {\enquote {\bibinfo {title} {Two-dimensional
  periodic frustrated ising models in a transverse field},}\ }\href {\doibase
  10.1103/PhysRevLett.84.4457} {\bibfield  {journal} {\bibinfo  {journal}
  {Phys. Rev. Lett.}\ }\textbf {\bibinfo {volume} {84}},\ \bibinfo {pages}
  {4457--4460} (\bibinfo {year} {2000}{\natexlab{b}})}\BibitemShut {NoStop}%
\bibitem [{\citenamefont {Samajdar}\ \emph {et~al.}(2021)\citenamefont
  {Samajdar}, \citenamefont {Ho}, \citenamefont {Pichler}, \citenamefont
  {Lukin},\ and\ \citenamefont {Sachdev}}]{pnas.2015785118}%
  \BibitemOpen
  \bibfield  {author} {\bibinfo {author} {\bibfnamefont {Rhine}\ \bibnamefont
  {Samajdar}}, \bibinfo {author} {\bibfnamefont {Wen~Wei}\ \bibnamefont {Ho}},
  \bibinfo {author} {\bibfnamefont {Hannes}\ \bibnamefont {Pichler}}, \bibinfo
  {author} {\bibfnamefont {Mikhail~D.}\ \bibnamefont {Lukin}}, \ and\ \bibinfo
  {author} {\bibfnamefont {Subir}\ \bibnamefont {Sachdev}},\ }\bibfield
  {title} {\enquote {\bibinfo {title} {Quantum phases of rydberg atoms on a
  kagome lattice},}\ }\href {\doibase 10.1073/pnas.2015785118} {\bibfield
  {journal} {\bibinfo  {journal} {Proceedings of the National Academy of
  Sciences}\ }\textbf {\bibinfo {volume} {118}},\ \bibinfo {pages}
  {e2015785118} (\bibinfo {year} {2021})},\ \Eprint
  {http://arxiv.org/abs/https://www.pnas.org/doi/pdf/10.1073/pnas.2015785118}
  {https://www.pnas.org/doi/pdf/10.1073/pnas.2015785118} \BibitemShut {NoStop}%
\bibitem [{\citenamefont {Shah}\ \emph {et~al.}(2025)\citenamefont {Shah},
  \citenamefont {Nambiar}, \citenamefont {Gorshkov},\ and\ \citenamefont
  {Galitski}}]{PhysRevX.15.011025}%
  \BibitemOpen
  \bibfield  {author} {\bibinfo {author} {\bibfnamefont {Jeet}\ \bibnamefont
  {Shah}}, \bibinfo {author} {\bibfnamefont {Gautam}\ \bibnamefont {Nambiar}},
  \bibinfo {author} {\bibfnamefont {Alexey~V.}\ \bibnamefont {Gorshkov}}, \
  and\ \bibinfo {author} {\bibfnamefont {Victor}\ \bibnamefont {Galitski}},\
  }\bibfield  {title} {\enquote {\bibinfo {title} {Quantum spin ice in
  three-dimensional rydberg atom arrays},}\ }\href {\doibase
  10.1103/PhysRevX.15.011025} {\bibfield  {journal} {\bibinfo  {journal} {Phys.
  Rev. X}\ }\textbf {\bibinfo {volume} {15}},\ \bibinfo {pages} {011025}
  (\bibinfo {year} {2025})}\BibitemShut {NoStop}%
\bibitem [{\citenamefont {Da~Liao}\ \emph {et~al.}(2021)\citenamefont
  {Da~Liao}, \citenamefont {Li}, \citenamefont {Yan}, \citenamefont {Wei},
  \citenamefont {Li}, \citenamefont {Qi},\ and\ \citenamefont
  {Meng}}]{PhysRevB.103.104416}%
  \BibitemOpen
  \bibfield  {author} {\bibinfo {author} {\bibfnamefont {Yuan}\ \bibnamefont
  {Da~Liao}}, \bibinfo {author} {\bibfnamefont {Han}\ \bibnamefont {Li}},
  \bibinfo {author} {\bibfnamefont {Zheng}\ \bibnamefont {Yan}}, \bibinfo
  {author} {\bibfnamefont {Hao-Tian}\ \bibnamefont {Wei}}, \bibinfo {author}
  {\bibfnamefont {Wei}\ \bibnamefont {Li}}, \bibinfo {author} {\bibfnamefont
  {Yang}\ \bibnamefont {Qi}}, \ and\ \bibinfo {author} {\bibfnamefont
  {Zi~Yang}\ \bibnamefont {Meng}},\ }\bibfield  {title} {\enquote {\bibinfo
  {title} {Phase diagram of the quantum ising model on a triangular lattice
  under external field},}\ }\href {\doibase 10.1103/PhysRevB.103.104416}
  {\bibfield  {journal} {\bibinfo  {journal} {Phys. Rev. B}\ }\textbf {\bibinfo
  {volume} {103}},\ \bibinfo {pages} {104416} (\bibinfo {year}
  {2021})}\BibitemShut {NoStop}%
\bibitem [{\citenamefont {Bombieri}\ \emph {et~al.}(2025)\citenamefont
  {Bombieri}, \citenamefont {Zache}, \citenamefont {Calliari}, \citenamefont
  {Lukin}, \citenamefont {Pichler},\ and\ \citenamefont
  {González-Cuadra}}]{bombieri2025deconfinedquantumcriticalitytriangular}%
  \BibitemOpen
  \bibfield  {author} {\bibinfo {author} {\bibfnamefont {Lisa}\ \bibnamefont
  {Bombieri}}, \bibinfo {author} {\bibfnamefont {Torsten~V.}\ \bibnamefont
  {Zache}}, \bibinfo {author} {\bibfnamefont {Gabriele}\ \bibnamefont
  {Calliari}}, \bibinfo {author} {\bibfnamefont {Mikhail~D.}\ \bibnamefont
  {Lukin}}, \bibinfo {author} {\bibfnamefont {Hannes}\ \bibnamefont {Pichler}},
  \ and\ \bibinfo {author} {\bibfnamefont {Daniel}\ \bibnamefont
  {González-Cuadra}},\ }\href {https://arxiv.org/abs/2508.08366} {\enquote
  {\bibinfo {title} {Deconfined quantum criticality on a triangular rydberg
  array},}\ } (\bibinfo {year} {2025}),\ \Eprint
  {http://arxiv.org/abs/2508.08366} {arXiv:2508.08366 [quant-ph]} \BibitemShut
  {NoStop}%
\bibitem [{\citenamefont {Ishizuka}\ \emph {et~al.}(2011)\citenamefont
  {Ishizuka}, \citenamefont {Motome}, \citenamefont {Furukawa},\ and\
  \citenamefont {Suzuki}}]{Ishizuka_2011}%
  \BibitemOpen
  \bibfield  {author} {\bibinfo {author} {\bibfnamefont {H}~\bibnamefont
  {Ishizuka}}, \bibinfo {author} {\bibfnamefont {Y}~\bibnamefont {Motome}},
  \bibinfo {author} {\bibfnamefont {N}~\bibnamefont {Furukawa}}, \ and\
  \bibinfo {author} {\bibfnamefont {S}~\bibnamefont {Suzuki}},\ }\bibfield
  {title} {\enquote {\bibinfo {title} {Quantum monte carlo study of the
  transverse-field ising model on a frustrated checkerboard lattice},}\ }\href
  {\doibase 10.1088/1742-6596/320/1/012054} {\bibfield  {journal} {\bibinfo
  {journal} {Journal of Physics: Conference Series}\ }\textbf {\bibinfo
  {volume} {320}},\ \bibinfo {pages} {012054} (\bibinfo {year}
  {2011})}\BibitemShut {NoStop}%
\bibitem [{\citenamefont {Merali}\ \emph {et~al.}(2024)\citenamefont {Merali},
  \citenamefont {De~Vlugt},\ and\ \citenamefont {Melko}}]{Melko_SSEforRydberg}%
  \BibitemOpen
  \bibfield  {author} {\bibinfo {author} {\bibfnamefont {Ejaaz}\ \bibnamefont
  {Merali}}, \bibinfo {author} {\bibfnamefont {Isaac J.~S.}\ \bibnamefont
  {De~Vlugt}}, \ and\ \bibinfo {author} {\bibfnamefont {Roger~G.}\ \bibnamefont
  {Melko}},\ }\bibfield  {title} {\enquote {\bibinfo {title} {Stochastic series
  expansion quantum monte carlo for rydberg arrays},}\ }\href {\doibase
  10.21468/SciPostPhysCore.7.2.016} {\bibfield  {journal} {\bibinfo  {journal}
  {SciPost Physics Core}\ }\textbf {\bibinfo {volume} {7}},\ \bibinfo {pages}
  {016} (\bibinfo {year} {2024})}\BibitemShut {NoStop}%
\bibitem [{\citenamefont {Patil}(2024)}]{QMC_rydberg}%
  \BibitemOpen
  \bibfield  {author} {\bibinfo {author} {\bibfnamefont {Pranay}\ \bibnamefont
  {Patil}},\ }\href {https://arxiv.org/abs/2309.00482} {\enquote {\bibinfo
  {title} {Quantum monte carlo simulations in the restricted hilbert space of
  rydberg atom arrays},}\ } (\bibinfo {year} {2024}),\ \Eprint
  {http://arxiv.org/abs/2309.00482} {arXiv:2309.00482 [cond-mat.str-el]}
  \BibitemShut {NoStop}%
\bibitem [{\citenamefont {{Evertz}}(2003)}]{qmc_loop}%
  \BibitemOpen
  \bibfield  {author} {\bibinfo {author} {\bibfnamefont {H.~G.}\ \bibnamefont
  {{Evertz}}},\ }\bibfield  {title} {\enquote {\bibinfo {title} {{The loop
  algorithm}},}\ }\href {\doibase 10.1080/0001873021000049195} {\bibfield
  {journal} {\bibinfo  {journal} {Advances in Physics}\ }\textbf {\bibinfo
  {volume} {52}},\ \bibinfo {pages} {1--66} (\bibinfo {year} {2003})},\ \Eprint
  {http://arxiv.org/abs/cond-mat/9707221} {arXiv:cond-mat/9707221
  [cond-mat.str-el]} \BibitemShut {NoStop}%
\bibitem [{\citenamefont {{Prokof'Ev}}\ \emph {et~al.}(1998)\citenamefont
  {{Prokof'Ev}}, \citenamefont {{Svistunov}},\ and\ \citenamefont
  {{Tupitsyn}}}]{worm1}%
  \BibitemOpen
  \bibfield  {author} {\bibinfo {author} {\bibfnamefont {N.~V.}\ \bibnamefont
  {{Prokof'Ev}}}, \bibinfo {author} {\bibfnamefont {B.~V.}\ \bibnamefont
  {{Svistunov}}}, \ and\ \bibinfo {author} {\bibfnamefont {I.~S.}\ \bibnamefont
  {{Tupitsyn}}},\ }\bibfield  {title} {\enquote {\bibinfo {title} {{Exact,
  complete, and universal continuous-time worldline Monte Carlo approach to the
  statistics of discrete quantum systems}},}\ }\href {\doibase
  10.1134/1.558661} {\bibfield  {journal} {\bibinfo  {journal} {Soviet Journal
  of Experimental and Theoretical Physics}\ }\textbf {\bibinfo {volume} {87}},\
  \bibinfo {pages} {310--321} (\bibinfo {year} {1998})},\ \Eprint
  {http://arxiv.org/abs/cond-mat/9703200} {arXiv:cond-mat/9703200 [cond-mat]}
  \BibitemShut {NoStop}%
\bibitem [{\citenamefont {{Alet}}\ \emph {et~al.}(2005)\citenamefont {{Alet}},
  \citenamefont {{Wessel}},\ and\ \citenamefont {{Troyer}}}]{worm2}%
  \BibitemOpen
  \bibfield  {author} {\bibinfo {author} {\bibfnamefont {Fabien}\ \bibnamefont
  {{Alet}}}, \bibinfo {author} {\bibfnamefont {Stefan}\ \bibnamefont
  {{Wessel}}}, \ and\ \bibinfo {author} {\bibfnamefont {Matthias}\ \bibnamefont
  {{Troyer}}},\ }\bibfield  {title} {\enquote {\bibinfo {title} {{Generalized
  directed loop method for quantum Monte Carlo simulations}},}\ }\href
  {\doibase 10.1103/PhysRevE.71.036706} {\bibfield  {journal} {\bibinfo
  {journal} {\pre}\ }\textbf {\bibinfo {volume} {71}},\ \bibinfo {eid} {036706}
  (\bibinfo {year} {2005})},\ \Eprint {http://arxiv.org/abs/cond-mat/0308495}
  {arXiv:cond-mat/0308495 [cond-mat.str-el]} \BibitemShut {NoStop}%
\bibitem [{\citenamefont {Sandvik}(1999)}]{SSE_Sandvik_1}%
  \BibitemOpen
  \bibfield  {author} {\bibinfo {author} {\bibfnamefont {Anders~W.}\
  \bibnamefont {Sandvik}},\ }\bibfield  {title} {\enquote {\bibinfo {title}
  {Stochastic series expansion method with operator-loop update},}\ }\href
  {\doibase 10.1103/PhysRevB.59.R14157} {\bibfield  {journal} {\bibinfo
  {journal} {Phys. Rev. B}\ }\textbf {\bibinfo {volume} {59}},\ \bibinfo
  {pages} {R14157--R14160} (\bibinfo {year} {1999})}\BibitemShut {NoStop}%
\bibitem [{sup()}]{sup}%
  \BibitemOpen
  \href@noop {} {}\bibinfo {note} {Teaching code:
  https://github.com/AGXFzhang/QTIM\_loop}\BibitemShut {NoStop}%
\bibitem [{\citenamefont {{Sandvik}}(1992)}]{measure}%
  \BibitemOpen
  \bibfield  {author} {\bibinfo {author} {\bibfnamefont {A.~W.}\ \bibnamefont
  {{Sandvik}}},\ }\bibfield  {title} {\enquote {\bibinfo {title} {{A
  generalization of Handscomb's quantum Monte Carlo scheme-application to the
  1D Hubbard model}},}\ }\href {\doibase 10.1088/0305-4470/25/13/017}
  {\bibfield  {journal} {\bibinfo  {journal} {Journal of Physics A Mathematical
  General}\ }\textbf {\bibinfo {volume} {25}},\ \bibinfo {pages} {3667--3682}
  (\bibinfo {year} {1992})}\BibitemShut {NoStop}%
\bibitem [{\citenamefont {Browaeys}\ and\ \citenamefont
  {Lahaye}(2020)}]{browaeys2020many}%
  \BibitemOpen
  \bibfield  {author} {\bibinfo {author} {\bibfnamefont {Antoine}\ \bibnamefont
  {Browaeys}}\ and\ \bibinfo {author} {\bibfnamefont {Thierry}\ \bibnamefont
  {Lahaye}},\ }\bibfield  {title} {\enquote {\bibinfo {title} {Many-body
  physics with individually controlled rydberg atoms},}\ }\href@noop {}
  {\bibfield  {journal} {\bibinfo  {journal} {Nat. Phys.}\ }\textbf {\bibinfo
  {volume} {16}},\ \bibinfo {pages} {132--142} (\bibinfo {year}
  {2020})}\BibitemShut {NoStop}%
\bibitem [{\citenamefont {{Liu}}\ \emph
  {et~al.}(2024{\natexlab{a}})\citenamefont {{Liu}}, \citenamefont {{Xu}},\
  and\ \citenamefont {{Zhang}}}]{RNG}%
  \BibitemOpen
  \bibfield  {author} {\bibinfo {author} {\bibfnamefont {Dong-Xu}\ \bibnamefont
  {{Liu}}}, \bibinfo {author} {\bibfnamefont {Wei}\ \bibnamefont {{Xu}}}, \
  and\ \bibinfo {author} {\bibfnamefont {Xue-Feng}\ \bibnamefont {{Zhang}}},\
  }\bibfield  {title} {\enquote {\bibinfo {title} {{Analysis of pseudo-random
  number generators in QMC-SSE method}},}\ }\href {\doibase
  10.1088/1674-1056/ad1e69} {\bibfield  {journal} {\bibinfo  {journal} {Chinese
  Physics B}\ }\textbf {\bibinfo {volume} {33}},\ \bibinfo {eid} {037509}
  (\bibinfo {year} {2024}{\natexlab{a}})},\ \Eprint
  {http://arxiv.org/abs/2403.06450} {arXiv:2403.06450 [cond-mat.str-el]}
  \BibitemShut {NoStop}%
\bibitem [{\citenamefont {{Moessner}}\ and\ \citenamefont
  {{Sondhi}}(2001{\natexlab{a}})}]{qtim_kagome1}%
  \BibitemOpen
  \bibfield  {author} {\bibinfo {author} {\bibfnamefont {R.}~\bibnamefont
  {{Moessner}}}\ and\ \bibinfo {author} {\bibfnamefont {S.~L.}\ \bibnamefont
  {{Sondhi}}},\ }\bibfield  {title} {\enquote {\bibinfo {title} {{Ising models
  of quantum frustration}},}\ }\href {\doibase 10.1103/PhysRevB.63.224401}
  {\bibfield  {journal} {\bibinfo  {journal} {\prb}\ }\textbf {\bibinfo
  {volume} {63}},\ \bibinfo {eid} {224401} (\bibinfo {year}
  {2001}{\natexlab{a}})},\ \Eprint {http://arxiv.org/abs/cond-mat/0011250}
  {arXiv:cond-mat/0011250 [cond-mat.stat-mech]} \BibitemShut {NoStop}%
\bibitem [{\citenamefont {{Moessner}}\ and\ \citenamefont
  {{Sondhi}}(2001{\natexlab{b}})}]{qtim_kagome2}%
  \BibitemOpen
  \bibfield  {author} {\bibinfo {author} {\bibfnamefont {R.}~\bibnamefont
  {{Moessner}}}\ and\ \bibinfo {author} {\bibfnamefont {S.~L.}\ \bibnamefont
  {{Sondhi}}},\ }\bibfield  {title} {\enquote {\bibinfo {title} {{Resonating
  Valence Bond Phase in the Triangular Lattice Quantum Dimer Model}},}\ }\href
  {\doibase 10.1103/PhysRevLett.86.1881} {\bibfield  {journal} {\bibinfo
  {journal} {\prl}\ }\textbf {\bibinfo {volume} {86}},\ \bibinfo {pages}
  {1881--1884} (\bibinfo {year} {2001}{\natexlab{b}})},\ \Eprint
  {http://arxiv.org/abs/cond-mat/0007378} {arXiv:cond-mat/0007378
  [cond-mat.str-el]} \BibitemShut {NoStop}%
\bibitem [{\citenamefont {{Zhang}}\ and\ \citenamefont
  {{Eggert}}(2013)}]{kagome_zhang1}%
  \BibitemOpen
  \bibfield  {author} {\bibinfo {author} {\bibfnamefont {Xue-Feng}\
  \bibnamefont {{Zhang}}}\ and\ \bibinfo {author} {\bibfnamefont {Sebastian}\
  \bibnamefont {{Eggert}}},\ }\bibfield  {title} {\enquote {\bibinfo {title}
  {{Chiral Edge States and Fractional Charge Separation in a System of
  Interacting Bosons on a Kagome Lattice}},}\ }\href {\doibase
  10.1103/PhysRevLett.111.147201} {\bibfield  {journal} {\bibinfo  {journal}
  {\prl}\ }\textbf {\bibinfo {volume} {111}},\ \bibinfo {eid} {147201}
  (\bibinfo {year} {2013})},\ \Eprint {http://arxiv.org/abs/1305.0003}
  {arXiv:1305.0003 [cond-mat.quant-gas]} \BibitemShut {NoStop}%
\bibitem [{\citenamefont {{Zhang}}\ \emph {et~al.}(2018)\citenamefont
  {{Zhang}}, \citenamefont {{He}}, \citenamefont {{Eggert}}, \citenamefont
  {{Moessner}},\ and\ \citenamefont {{Pollmann}}}]{kagome_zhang2}%
  \BibitemOpen
  \bibfield  {author} {\bibinfo {author} {\bibfnamefont {Xue-Feng}\
  \bibnamefont {{Zhang}}}, \bibinfo {author} {\bibfnamefont {Yin-Chen}\
  \bibnamefont {{He}}}, \bibinfo {author} {\bibfnamefont {Sebastian}\
  \bibnamefont {{Eggert}}}, \bibinfo {author} {\bibfnamefont {Roderich}\
  \bibnamefont {{Moessner}}}, \ and\ \bibinfo {author} {\bibfnamefont {Frank}\
  \bibnamefont {{Pollmann}}},\ }\bibfield  {title} {\enquote {\bibinfo {title}
  {{Continuous Easy-Plane Deconfined Phase Transition on the Kagome
  Lattice}},}\ }\href {\doibase 10.1103/PhysRevLett.120.115702} {\bibfield
  {journal} {\bibinfo  {journal} {\prl}\ }\textbf {\bibinfo {volume} {120}},\
  \bibinfo {eid} {115702} (\bibinfo {year} {2018})},\ \Eprint
  {http://arxiv.org/abs/1706.05414} {arXiv:1706.05414 [cond-mat.str-el]}
  \BibitemShut {NoStop}%
\bibitem [{\citenamefont {{Liu}}\ \emph
  {et~al.}(2024{\natexlab{b}})\citenamefont {{Liu}}, \citenamefont {{Xiong}},
  \citenamefont {{Xu}},\ and\ \citenamefont {{Zhang}}}]{kagome_zhang3}%
  \BibitemOpen
  \bibfield  {author} {\bibinfo {author} {\bibfnamefont {Dong-Xu}\ \bibnamefont
  {{Liu}}}, \bibinfo {author} {\bibfnamefont {Zijian}\ \bibnamefont {{Xiong}}},
  \bibinfo {author} {\bibfnamefont {Yining}\ \bibnamefont {{Xu}}}, \ and\
  \bibinfo {author} {\bibfnamefont {Xue-Feng}\ \bibnamefont {{Zhang}}},\
  }\bibfield  {title} {\enquote {\bibinfo {title} {{Deconfined quantum phase
  transition on the kagome lattice: Distinct velocities of spinon and string
  excitations}},}\ }\href {\doibase 10.1103/PhysRevB.109.L140404} {\bibfield
  {journal} {\bibinfo  {journal} {\prb}\ }\textbf {\bibinfo {volume} {109}},\
  \bibinfo {eid} {L140404} (\bibinfo {year} {2024}{\natexlab{b}})},\ \Eprint
  {http://arxiv.org/abs/2301.12864} {arXiv:2301.12864 [cond-mat.str-el]}
  \BibitemShut {NoStop}%
\bibitem [{\citenamefont {Xu}\ and\ \citenamefont
  {Zhang}(2025)}]{xu2025geometricbreakingquantumstrings}%
  \BibitemOpen
  \bibfield  {author} {\bibinfo {author} {\bibfnamefont {Wei}\ \bibnamefont
  {Xu}}\ and\ \bibinfo {author} {\bibfnamefont {Xue-Feng}\ \bibnamefont
  {Zhang}},\ }\href {https://arxiv.org/abs/2410.21135} {\enquote {\bibinfo
  {title} {Geometric breaking of quantum strings in kagome rydberg atom
  array},}\ } (\bibinfo {year} {2025}),\ \Eprint
  {http://arxiv.org/abs/2410.21135} {arXiv:2410.21135 [cond-mat.quant-gas]}
  \BibitemShut {NoStop}%
\bibitem [{\citenamefont {Liu}\ \emph {et~al.}(2020{\natexlab{b}})\citenamefont
  {Liu}, \citenamefont {Huang},\ and\ \citenamefont {Chen}}]{chunjiong}%
  \BibitemOpen
  \bibfield  {author} {\bibinfo {author} {\bibfnamefont {Changle}\ \bibnamefont
  {Liu}}, \bibinfo {author} {\bibfnamefont {Chun-Jiong}\ \bibnamefont {Huang}},
  \ and\ \bibinfo {author} {\bibfnamefont {Gang}\ \bibnamefont {Chen}},\
  }\bibfield  {title} {\enquote {\bibinfo {title} {Intrinsic quantum ising
  model on a triangular lattice magnet
  $\mathrm{Tm}\mathrm{Mg}\mathrm{Ga}{\mathrm{o}}_{4}$},}\ }\href {\doibase
  10.1103/PhysRevResearch.2.043013} {\bibfield  {journal} {\bibinfo  {journal}
  {Phys. Rev. Res.}\ }\textbf {\bibinfo {volume} {2}},\ \bibinfo {pages}
  {043013} (\bibinfo {year} {2020}{\natexlab{b}})}\BibitemShut {NoStop}%
\bibitem [{\citenamefont {Wang}\ \emph {et~al.}(2025)\citenamefont {Wang},
  \citenamefont {Wang}, \citenamefont {Ding}, \citenamefont {Mao},\ and\
  \citenamefont {Yan}}]{Yan_1}%
  \BibitemOpen
  \bibfield  {author} {\bibinfo {author} {\bibfnamefont {Zhe}\ \bibnamefont
  {Wang}}, \bibinfo {author} {\bibfnamefont {Zhiyan}\ \bibnamefont {Wang}},
  \bibinfo {author} {\bibfnamefont {Yi-Ming}\ \bibnamefont {Ding}}, \bibinfo
  {author} {\bibfnamefont {Bin-Bin}\ \bibnamefont {Mao}}, \ and\ \bibinfo
  {author} {\bibfnamefont {Zheng}\ \bibnamefont {Yan}},\ }\bibfield  {title}
  {\enquote {\bibinfo {title} {Bipartite reweight-annealing algorithm of
  quantum monte carlo to extract large-scale data of entanglement entropy and
  its derivative},}\ }\href {\doibase 10.1038/s41467-025-61084-7} {\bibfield
  {journal} {\bibinfo  {journal} {Nature Communications}\ }\textbf {\bibinfo
  {volume} {16}},\ \bibinfo {pages} {5880} (\bibinfo {year}
  {2025})}\BibitemShut {NoStop}%
\bibitem [{\citenamefont {Ding}\ \emph
  {et~al.}(2025{\natexlab{a}})\citenamefont {Ding}, \citenamefont {Tang},
  \citenamefont {Wang}, \citenamefont {Wang}, \citenamefont {Mao},\ and\
  \citenamefont {Yan}}]{Yan_2}%
  \BibitemOpen
  \bibfield  {author} {\bibinfo {author} {\bibfnamefont {Yi-Ming}\ \bibnamefont
  {Ding}}, \bibinfo {author} {\bibfnamefont {Yin}\ \bibnamefont {Tang}},
  \bibinfo {author} {\bibfnamefont {Zhe}\ \bibnamefont {Wang}}, \bibinfo
  {author} {\bibfnamefont {Zhiyan}\ \bibnamefont {Wang}}, \bibinfo {author}
  {\bibfnamefont {Bin-Bin}\ \bibnamefont {Mao}}, \ and\ \bibinfo {author}
  {\bibfnamefont {Zheng}\ \bibnamefont {Yan}},\ }\bibfield  {title} {\enquote
  {\bibinfo {title} {Tracking the variation of entanglement r\'enyi negativity:
  A quantum monte carlo study},}\ }\href {\doibase
  10.1103/PhysRevB.111.L241108} {\bibfield  {journal} {\bibinfo  {journal}
  {Phys. Rev. B}\ }\textbf {\bibinfo {volume} {111}},\ \bibinfo {pages}
  {L241108} (\bibinfo {year} {2025}{\natexlab{a}})}\BibitemShut {NoStop}%
\bibitem [{\citenamefont {Ding}\ \emph
  {et~al.}(2025{\natexlab{b}})\citenamefont {Ding}, \citenamefont {Wang},\ and\
  \citenamefont {Yan}}]{Yan_3}%
  \BibitemOpen
  \bibfield  {author} {\bibinfo {author} {\bibfnamefont {Yi-Ming}\ \bibnamefont
  {Ding}}, \bibinfo {author} {\bibfnamefont {Zhe}\ \bibnamefont {Wang}}, \ and\
  \bibinfo {author} {\bibfnamefont {Zheng}\ \bibnamefont {Yan}},\ }\bibfield
  {title} {\enquote {\bibinfo {title} {Evaluating many-body stabilizer r\'enyi
  entropy by sampling reduced pauli strings: Singularities, volume law, and
  nonlocal magic},}\ }\href {\doibase 10.1103/pyzr-jmvw} {\bibfield  {journal}
  {\bibinfo  {journal} {PRX Quantum}\ }\textbf {\bibinfo {volume} {6}},\
  \bibinfo {pages} {030328} (\bibinfo {year} {2025}{\natexlab{b}})}\BibitemShut
  {NoStop}%
\bibitem [{\citenamefont {Mao}\ \emph {et~al.}(2025)\citenamefont {Mao},
  \citenamefont {Ding}, \citenamefont {Wang}, \citenamefont {Hu},\ and\
  \citenamefont {Yan}}]{Yan_4}%
  \BibitemOpen
  \bibfield  {author} {\bibinfo {author} {\bibfnamefont {Bin-Bin}\ \bibnamefont
  {Mao}}, \bibinfo {author} {\bibfnamefont {Yi-Ming}\ \bibnamefont {Ding}},
  \bibinfo {author} {\bibfnamefont {Zhe}\ \bibnamefont {Wang}}, \bibinfo
  {author} {\bibfnamefont {Shijie}\ \bibnamefont {Hu}}, \ and\ \bibinfo
  {author} {\bibfnamefont {Zheng}\ \bibnamefont {Yan}},\ }\bibfield  {title}
  {\enquote {\bibinfo {title} {Sampling reduced density matrix to extract fine
  levels of entanglement spectrum and restore entanglement hamiltonian},}\
  }\href {\doibase 10.1038/s41467-025-58058-0} {\bibfield  {journal} {\bibinfo
  {journal} {Nature Communications}\ }\textbf {\bibinfo {volume} {16}},\
  \bibinfo {pages} {2880} (\bibinfo {year} {2025})}\BibitemShut {NoStop}%
\bibitem [{\citenamefont {Ding}\ \emph {et~al.}(2024)\citenamefont {Ding},
  \citenamefont {Sun}, \citenamefont {Ma}, \citenamefont {Pan}, \citenamefont
  {Cheng},\ and\ \citenamefont {Yan}}]{Yan_5}%
  \BibitemOpen
  \bibfield  {author} {\bibinfo {author} {\bibfnamefont {Yi-Ming}\ \bibnamefont
  {Ding}}, \bibinfo {author} {\bibfnamefont {Jun-Song}\ \bibnamefont {Sun}},
  \bibinfo {author} {\bibfnamefont {Nvsen}\ \bibnamefont {Ma}}, \bibinfo
  {author} {\bibfnamefont {Gaopei}\ \bibnamefont {Pan}}, \bibinfo {author}
  {\bibfnamefont {Chen}\ \bibnamefont {Cheng}}, \ and\ \bibinfo {author}
  {\bibfnamefont {Zheng}\ \bibnamefont {Yan}},\ }\bibfield  {title} {\enquote
  {\bibinfo {title} {Reweight-annealing method for evaluating the partition
  function via quantum monte carlo calculations},}\ }\href {\doibase
  10.1103/PhysRevB.110.165152} {\bibfield  {journal} {\bibinfo  {journal}
  {Phys. Rev. B}\ }\textbf {\bibinfo {volume} {110}},\ \bibinfo {pages}
  {165152} (\bibinfo {year} {2024})}\BibitemShut {NoStop}%
\bibitem [{\citenamefont {Lee}\ \emph {et~al.}(1984)\citenamefont {Lee},
  \citenamefont {Joannopoulos},\ and\ \citenamefont
  {Negele}}]{PhysRevB.30.1599}%
  \BibitemOpen
  \bibfield  {author} {\bibinfo {author} {\bibfnamefont {D.~H.}\ \bibnamefont
  {Lee}}, \bibinfo {author} {\bibfnamefont {J.~D.}\ \bibnamefont
  {Joannopoulos}}, \ and\ \bibinfo {author} {\bibfnamefont {J.~W.}\
  \bibnamefont {Negele}},\ }\bibfield  {title} {\enquote {\bibinfo {title}
  {Monte carlo solution of antiferromagnetic quantum heisenberg spin
  systems},}\ }\href {\doibase 10.1103/PhysRevB.30.1599} {\bibfield  {journal}
  {\bibinfo  {journal} {Phys. Rev. B}\ }\textbf {\bibinfo {volume} {30}},\
  \bibinfo {pages} {1599--1602} (\bibinfo {year} {1984})}\BibitemShut {NoStop}%
\end{thebibliography}%
\end{document}